\documentclass[reprint,aps,pra,showkeys,showpacs]{revtex4-2}  % like in a journal
\usepackage{epsfig}
\usepackage{xcolor}   % To add color to text
\usepackage{amsmath}
\usepackage{epstopdf}

\begin{document}

\title{Beyond mean-field effects in Josephson oscillations and self-trapping of Bose-Einstein condensates in two-dimensional dual-core traps}

\author{Sherzod R. Otajonov$^{1,2}$}
\author{Fatkhulla Kh. Abdullaev$^{1,3}$}
\author{Akbar Shermaxmatov$^{1}$}
\address{$^1$ Uzbekistan Academy of Sciences S. A. Azimov Physical-Technical Institute, Chingiz Aytmatov Str. 2-B, 100084, Tashkent, Uzbekistan}
\address{$^2$ National University of Uzbekistan, Department of Theoretical Physics, 100174, Tashkent, Uzbekistan}
\address{$^3$ Institute of Theoretical Physics, National University of Uzbekistan, 100174, Tashkent, Uzbekistan}
%\address{$^4$ "TIIAME" National Research University, Tashkent 100000, Uzbekistan}

%
\begin{abstract}
We study a binary Bose gas in a symmetric dual-core, pancake-shaped trap, modelled by two linearly coupled two-dimensional Gross-Pitaevskii equations with Lee-Huang-Yang corrections. Two different cases are considered. First, we consider a spatially uniform condensate, where we identify the domains of parameters for macroscopic quantum tunnelling, self-trapping and localisation revivals. The analytical formulas for the Josephson frequencies in the zero- and $\pi$-phase modes are derived. As the total atom number varies, the system displays a rich bifurcation structure. In the zero-phase, two successive pitchfork bifurcations generate bistability and hysteresis, while the $\pi$-phase exhibits a single pitchfork bifurcation.

The second case is when the quantum droplets are in a dual-core trap. Analytical predictions for the oscillation frequencies are derived via a variational approach for the coupled dynamics of quantum droplets, and direct numerical simulations validate the results. We identify critical values of the linear coupling that separate Josephson and self-trapped regimes as the particle number changes. We also found the Andreev-Bashkin superfluid drag effect in numerical simulations of the droplet-droplet interactions in the two-core geometry.
\end{abstract}
\maketitle

\section{Introduction}
\label{intro}
Recently, substantial attention has been devoted to beyond-mean-field effects in ultracold gases. These effects arise from quantum fluctuations (QFs), which can dynamically stabilise an otherwise collapsing Bose-Einstein condensate (BEC). For example, in binary Bose mixtures with weak residual attractive mean-field interactions, QFs can arrest collapse~\cite{Petrov, PA}. In three dimensions, the QF contribution is described by the Lee-Huang-Yang (LHY) correction and takes the form of a repulsive term $\propto n^{5/2}$, where $n$ is the condensate density. The balance between residual mean-field attraction and QF-induced repulsion produces new localised states, quantum droplets which have been observed in binary BECs~\cite{exp1,exp2} and in single-component dipolar BECs~\cite{dipol}.

Beyond the study of QFs-induced stationary states, it is important to analyse non-stationary coherent processes, notably the Josephson effect in quantum gases. In bosonic Josephson junctions, phenomena such as macroscopic quantum tunnelling, self-trapping, bifurcations, and homoclinic structures are predicted and have been explored both theoretically and experimentally~\cite{Albiez1, Albiez2, Albiez3, Smerzi1997, TM, Chen2020, Tilek2020, Sahem2002, Sinha2019, Mondal2022}. In addition, spontaneous symmetry breaking has been reported in one- and two-dimensional BECs coupled through macroscopic quantum tunneling~\cite{Chen2016,Chen2023}.

Because macroscopic quantum tunnelling and self-trapping are sensitive to the nonlinear terms in the extended Gross-Pitaevskii equation, QFs are expected to significantly affect dynamical processes in atomic Josephson junctions. To date, only a few studies have examined beyond-mean-field effects in such junctions. 
In the work~\cite{Sal2020}, a two-mode model for one-component stable condensate with repulsive mean-field interactions and the LHY correction considered as a small perturbation is investigated.

The extended Gross-Pitaevskii equation with an LHY term in a double-well potential has also been investigated: for quasi-one-dimensional Bose-Bose mixtures~\cite{Abd1, Pol} and for elongated geometries~\cite{Abd2}. Stationary states in a four-well potential for a two-dimensional BEC with LHY corrections were obtained in~\cite{Mihal2023}. The Josephson effect between droplets has been explored for two interacting three-dimensional droplets in a symmetric Bose-Bose mixture~\cite{Pylak22}, where the interplay between Josephson oscillations and droplet motion was treated under a weak-overlap approximation. Josephson dynamics of quantum droplets in a quasi-one-dimensional two-core trap were analysed in~\cite{AGS25}.

Understanding Josephson oscillations between droplets is also important for measuring the superfluid fraction in dipolar-BEC supersolids~\cite{nature1,nature2}. Here, we study macroscopic quantum tunnelling and self-trapping of matter waves in a two-dimensional two-core trap while accounting for beyond mean-field effects. We consider two types of states: homogeneous condensates and quantum droplets, and we analyse Josephson oscillations and self-trapping in both settings.

The paper is organised as follows. Section~\ref{sec:model} introduces the model. Sections~\ref{sec:dimer} and~\ref{sec:bifurcation} analyse the dimer model and its bifurcations. Droplet interactions are investigated in Section~\ref{sec:QDinteraction}. The Josephson effect for moving droplets is discussed in Section~\ref{sec:movingQDs}.

\section{The model}
\label{sec:model}

We consider the dynamics of a two-component BEC loaded in the dual-core cigar-type trap, when the beyond mean-field effects, describing quantum fluctuations, are taken into account. The case when the wave functions of each component inside each core are equal is considered. In this case, the system is described by two coupled modified Gross-Pitaevskii equations of the form~\cite{Petrov, PA, Malomed19}:
\begin{eqnarray}
& i\hbar\Psi_{1,T}=-\frac{\hbar^2}{2m_0}\nabla^2\Psi_1  +V(\rho)\Psi_1 + \nonumber \\
& \bar g|\Psi_1|^2\Psi_1 \ln(\frac{|\Psi_1|^2}{\sqrt{e}n_0})-\bar K\Psi_2,\nonumber \\
& i\hbar\Psi_{2,T}=-\frac{\hbar^2}{2m_0}\nabla^2\Psi_2  +V(\rho)\Psi_2 + 
\nonumber \\
& \bar g|\Psi_2|^2\Psi_2 \ln(\frac{|\Psi_2|^2}{\sqrt{e}n_0}) -\bar K\Psi_1 .
\end{eqnarray}
where $m_0$ is atomic mass, $\bar g=8\pi\hbar\omega_{\perp}/\ln^{2}(a_{12}/a)$ with $\omega_{\perp}$ denoting the transverse confinement frequency, and where $a_{11}=a_{22}\equiv a$ and $a_{12}$ represent the intra- and inter-species $s$-wave scattering lengths, respectively. 
The scattering lengths here are defined in 2D, and their relation to the corresponding 3D scattering lengths is given by
$$
a_{ij}^{(2D)} = \left(\frac{4\pi}{B}\right)^{1/2} l_0 
\exp\!\left(-\gamma - \sqrt{\tfrac{\pi}{2}}\,\frac{l_0}{a_{ij}^{(3D)}}\right),
$$
where $B=0.915$, $l_0 = (\hbar/m_0 \omega_{\perp})^{1/2}$ is the harmonic oscillator length, and $\gamma$ is the Euler-Mascheroni constant (see Ref.~\cite{PA}). The scattering length is expressed as $a_{ij}$ with $a_{12}=a_{21}$, $i, j=1$ and $2$.

The system in dimensionless form is:
\begin{eqnarray}
&iu_t + \frac{1}{2}\nabla^2 u - g\ln(|u|^2) |u|^2 u + K v=0, 
\nonumber \\
&iv_t +\frac{1}{2}\nabla^2 v -g\ln(|v|^2)|v|^2 v + K u=0,
\label{eq:coupGPE}
\end{eqnarray}
where
$$
g=\frac{8\pi\sqrt{e}n_0}{\ln^2(a_{12}/a)}, \ n_0 = \frac{e^{-2\gamma -3/2}}{2\pi}\frac{\ln(a_{12}/a)}{a a_{12}}, \ K=\frac{\bar K}{\hbar \omega_{\perp}}.
$$

Related work that incorporates quantum fluctuations in a setting closely aligned with our model has been carried out in one dimension, see Ref.~\cite{Malomed19}. In Ref.~\cite{Sal2020}, the role of dimensionality in atomic Josephson junctions with quantum fluctuations is analysed. The authors employ a two-mode approximation for a two-dimensional Bose-Einstein condensate with repulsive mean-field interactions, supplemented by a weak (perturbative) LHY correction. Owing to the logarithmic renormalisation of the effective coupling in 2D, quantum fluctuations enhance the interaction energy, leading to an upshift of the Josephson oscillation frequency and a lowering of the self-trapping threshold (i.e., facilitating the onset of macroscopic quantum self-trapping).

\section{Dimer limit}
\label{sec:dimer}

Let us consider the case of homogeneous in the space fields i.e.$u(x,t), v(x,t)$.

Then we have the LHY dimer equations 
\begin{eqnarray}
iu_t =g|u|^2 u \ln(|u|^2) - K v,\nonumber\\
iv_t = g|v|^2v\ln(|v|^2) - K u.
\label{eq:LHYdimer}
\end{eqnarray}
Representing fields in the form
$$
u=A_1 e^{i\phi_1}, \  v = A_2 e^{i\phi_2}.
$$
and introducing the relative variables $Z$ is the population imbalance and the relative phase $\theta$ according to:
$$
Z=\frac{N_2-N_1}{N}, \ \theta =\phi_2 -\phi_1, N_i=A_i^2, N = N_1 + N_2,
$$
we obtain the system of equations:
\begin{eqnarray} \label{eq:dimerZtThetat}
&Z_t=2 K \sqrt{1-Z^2}\sin(\theta), \\
\nonumber
&\theta_t =\frac{gN}{2}[(1-Z) \ln(\frac{N(1-Z)}{2})-\\
\nonumber
&-(1+Z) \ln(\frac{N(1+Z)}{2})] -\frac{2 K Z}{\sqrt{1-Z^2}}\cos(\theta).
\end{eqnarray}

The frequency of the Josephson oscillations:

i) zero-phase mode ($\theta=0$):
\begin{equation} \label{eq:Jfzero}
\omega_J^{(0)} = 2K \sqrt{1+\frac{gN}{2K} \ln\left(\frac{e N}{2} \right)},
\end{equation}

ii) $\pi$-phase mode($\theta=\pi$):
\begin{equation} \label{eq:Jfpi}
\omega_J^{(\pi)} = 2K \sqrt{1-\frac{gN}{2K} \ln\left(\frac{e N}{2} \right)}.
\end{equation}
Considering $Z$ and $\theta$ as canonical variables, we can write the Hamilton equations:
\begin{equation*}
    Z_t=-\frac{\partial H}{\partial \theta},   \theta_t=\frac{\partial H}{\partial Z}.
\end{equation*}
The Hamiltonian for the system (\ref{eq:dimerZtThetat}) is:
\begin{eqnarray} \label{HamiltonianPW}
H&=&2 K \sqrt{1-Z^2}\cos\theta -\frac{g N}{4} [ 2 Z^2 \ln(\frac{N}{2 \sqrt{e}}) \nonumber\\
&+& (1-Z)^2 \ln(1-Z) + (1+Z)^2 \ln(1+Z) ].
\end{eqnarray}

The condition for the transition from the macroscopic quantum tunelling regime to the self-trapping mode is:
\begin{equation}
    H(Z_0,\theta_0)>\mp2K,
\end{equation}
where $Z_0$ and $\theta_0$ are the initial imbalance and phase, and we obtain the following expression for the critical value
\begin{eqnarray} \label{CV}
   & K_{\mathrm{cr}}=gN(2Z_0^2\ln{\frac{N}{2\sqrt{e}}}+(1+Z_0)^2\ln{(1+Z_0)}+\nonumber\\
&(1-Z_0)^2\ln{(1-Z_0})) \times \nonumber \\
& (8(\sqrt{1-Z_0^2}\cos{\theta_0}\pm1))^{-1}.
\end{eqnarray}
% Figure 1
\begin{figure}[htbp]
  \centerline{ \includegraphics[width=4.4cm]{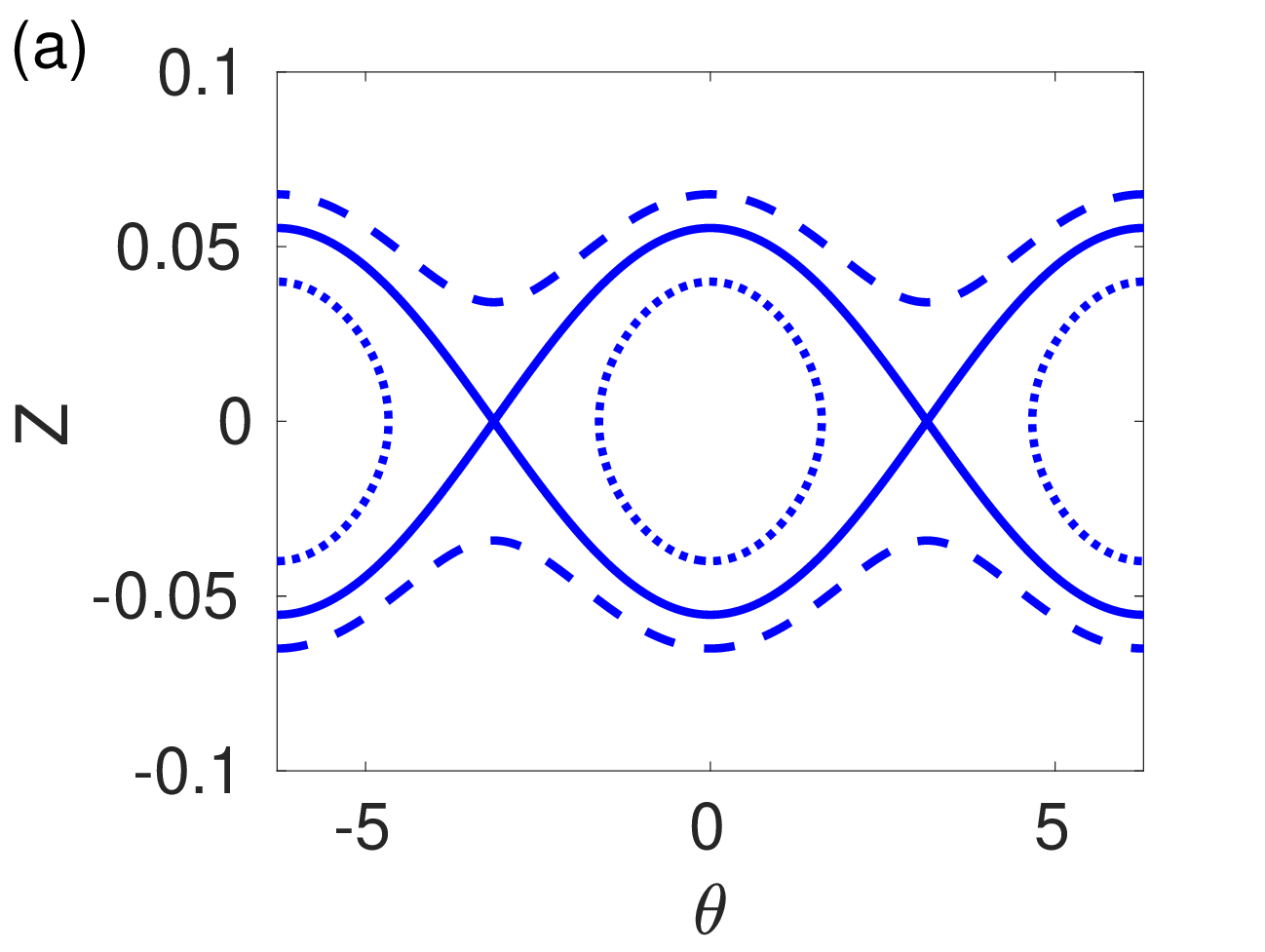}
  \includegraphics[width=4.4cm]{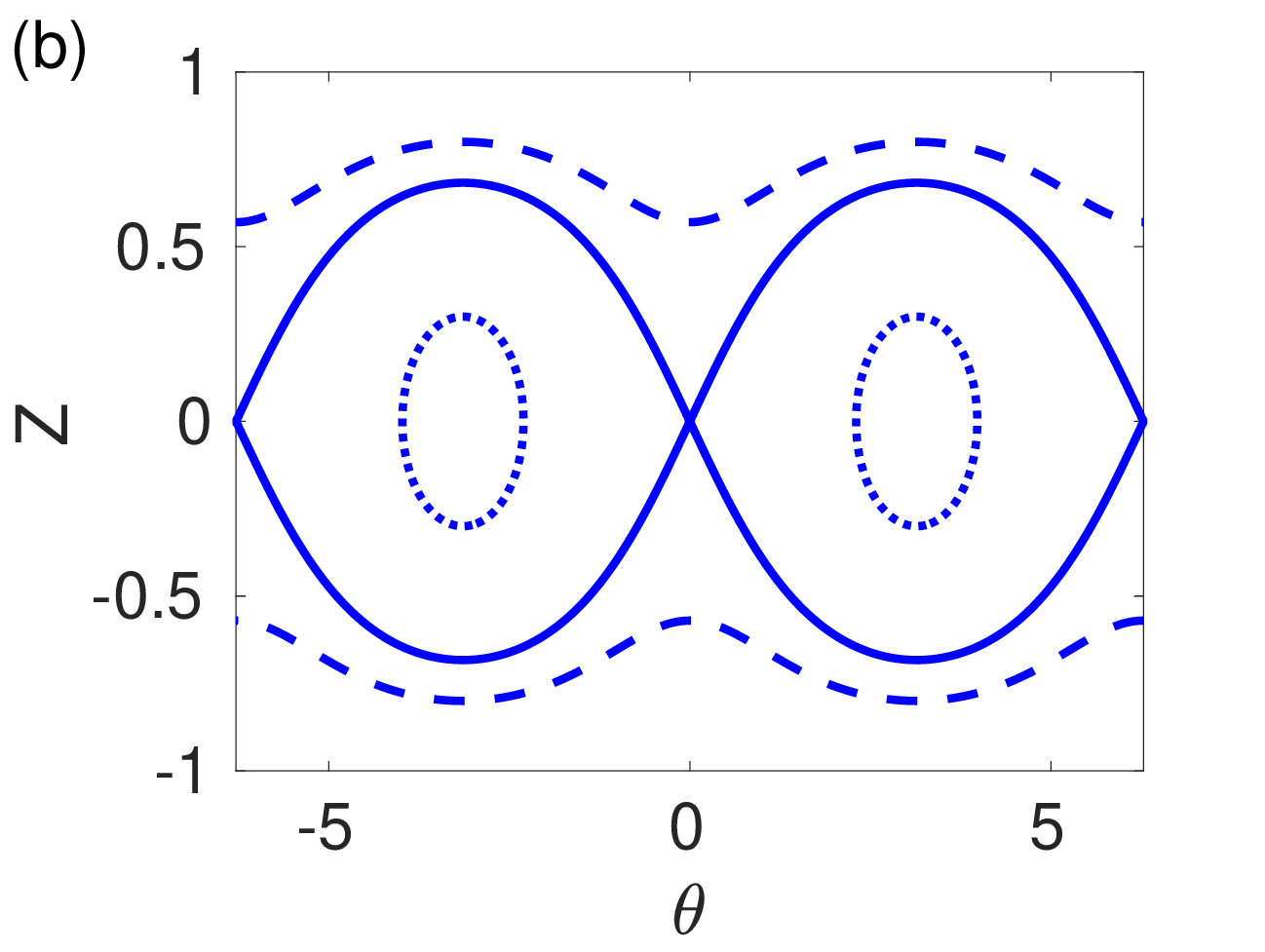} }
  \includegraphics[width=4.4cm]{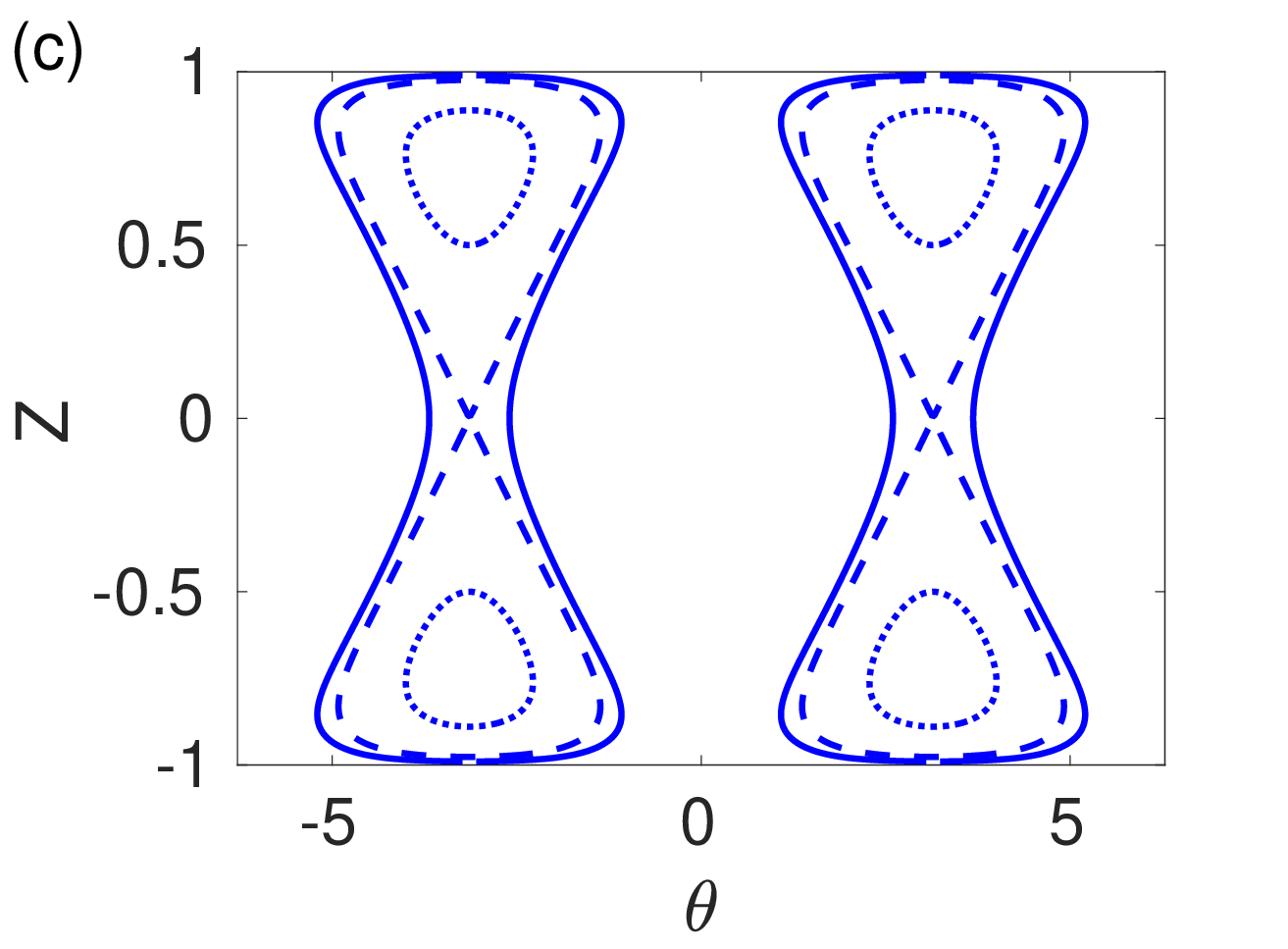}
\caption{These phase portraits are presented for different initial imbalances and different phase modes: $Z_0=0.04$ (dotted line), $0.0553$ (corresponds to a solid line and a critical value), $0.065$ (dashed line), $\theta_0=0$, $N=10$ (a), $Z_0=0.3$ (dotted line), $0.6829$ (corresponds to a solid line and a critical value), $0.8$ (dashed line), $\theta_0=\pi$, $N=0.6$ (b), $Z_0=0.5$ (dotted line), $0.99$ (solid line), $0.004$ (dashed line), $\theta_0=\pi$, $N=0.85$ (c). Other parameters $K=0.01$ and $g=1$.}
\label{fig1}
\end{figure}
Figure~\ref{fig1} illustrates the phase portraits derived from the Hamiltonian function given by Eq.(\ref{HamiltonianPW}). In Figure \ref{fig1} (a), the zero-phase mode is analysed for various initial values of the population imbalance $Z_0$. When the initial imbalance is small, e.g., $Z_0 = 0.04$, Josephson oscillations are observed between the two condensates (dotted line). As $Z_0$ increases, the character of the dynamics changes, leading to a regime in which atoms remain predominantly localised, a signature of self-trapping behaviour (dashed line). The critical value separating the Josephson and self-trapping regimes is $Z_{0c} = 0.553$ (solid line), consistent with Eq.~(\ref{CV}).

Figure~\ref{fig1}(b) presents the corresponding behaviour in the $\pi$-phase regime. A similar transition from Josephson oscillations to self-trapping is observed; however, this transition occurs only when the number of atoms is sufficiently small. In this case, the critical imbalance is $Z_{0c} = 0.6829$, which also agrees with the prediction of Eq.~(\ref{CV}).

Figure \ref{fig1} (c) illustrates the case of $N=0.85$. For $Z_0=0.004$, a localisation-like mode appears, but it differs from the self-trapping mode. This regime is known as localisation-revival, where the relative imbalance $Z(t)$ oscillates almost between $0$ and $1$. At $Z_0=0.5$, a self-trapping mode occurs: the relative phase does not grow indefinitely, but instead oscillates harmonically around the $\pi$-phase. For $Z_0=0.99$, Josephson oscillations are observed; however, these oscillations deviate from being perfectly harmonic.

% Figure 2
\begin{figure}[htbp]
  \centering
\includegraphics[width=4cm]{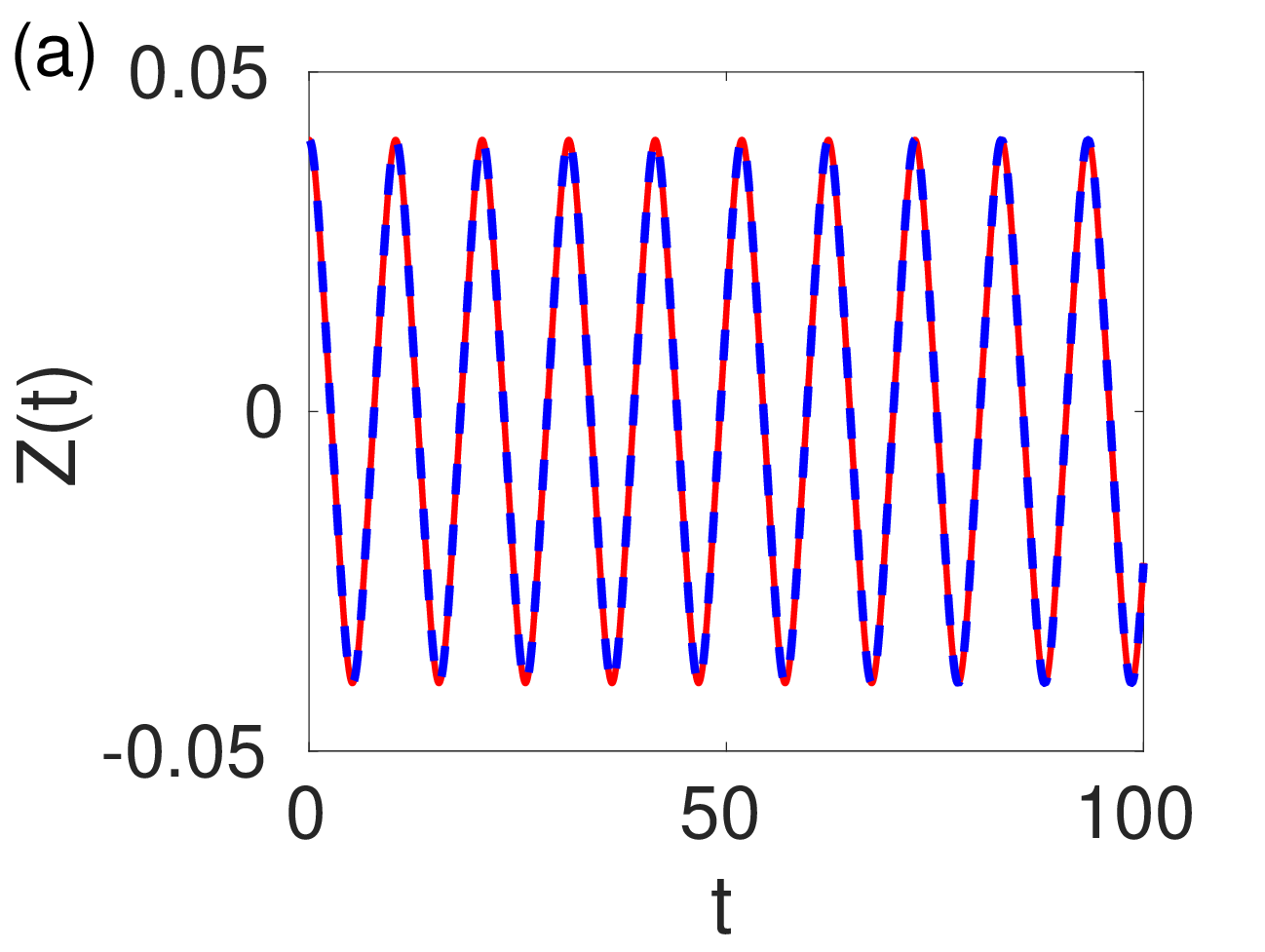}
\includegraphics[width=4cm]{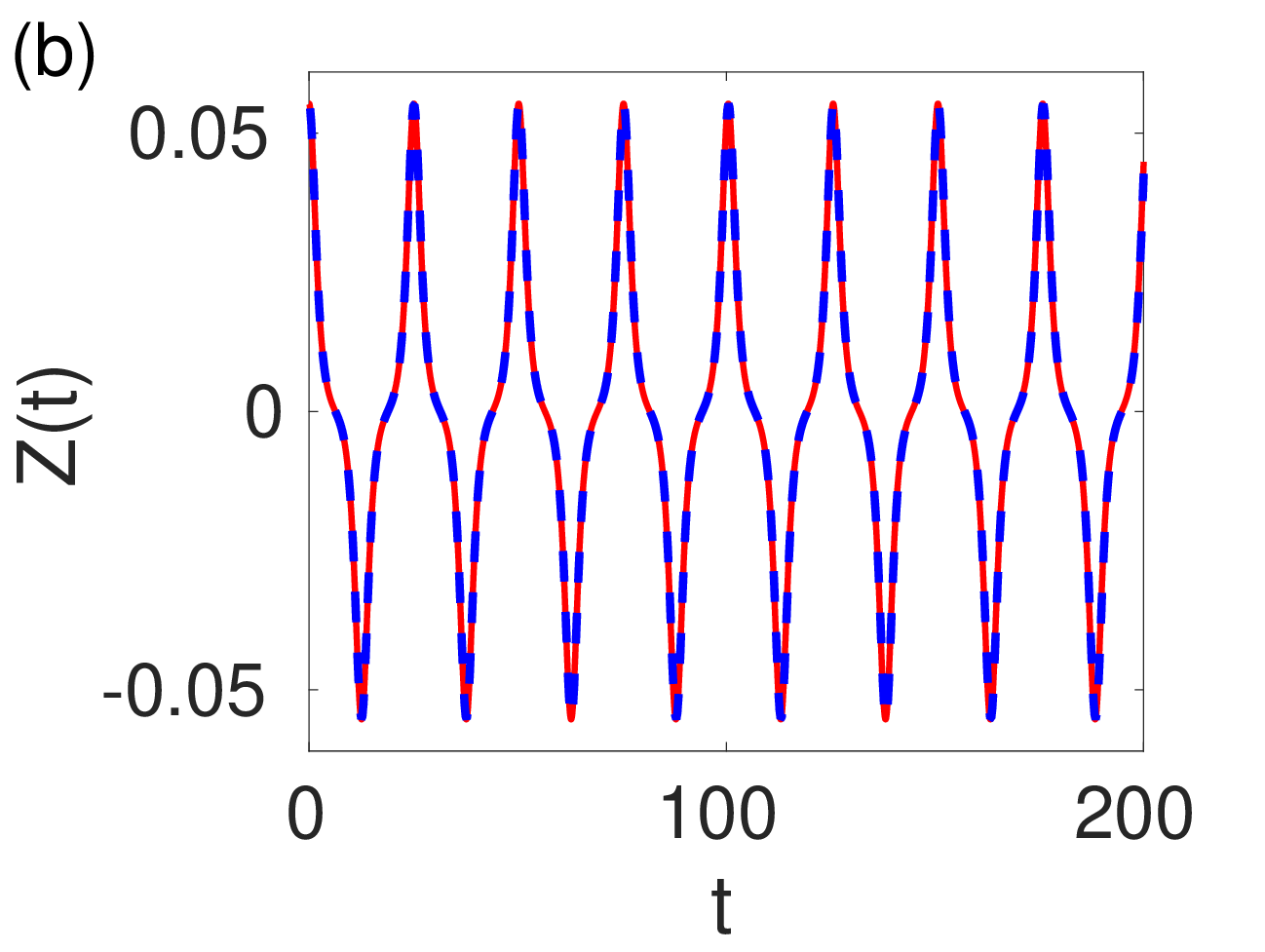}
\includegraphics[width=4cm]{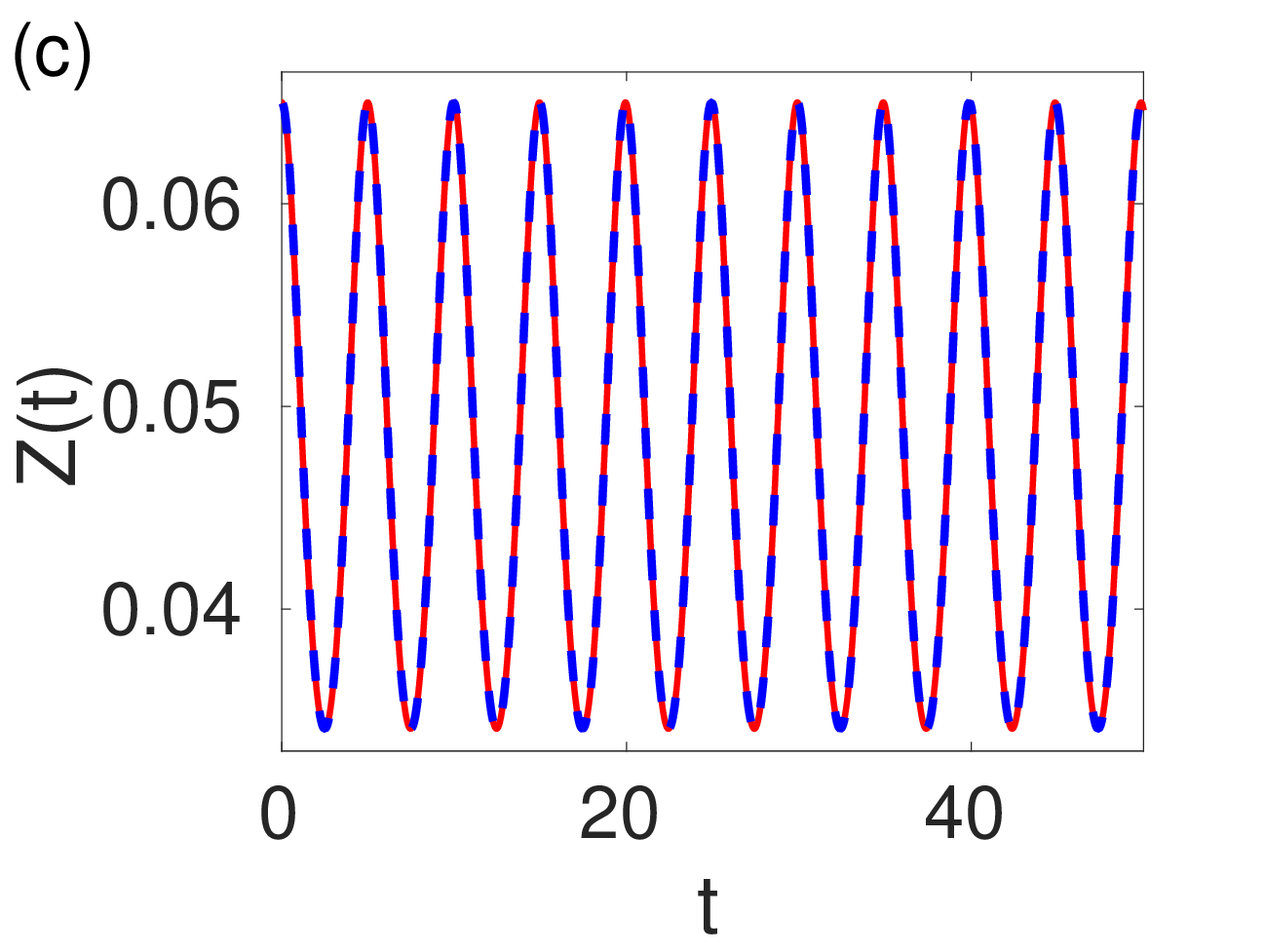}
\includegraphics[width=4cm]{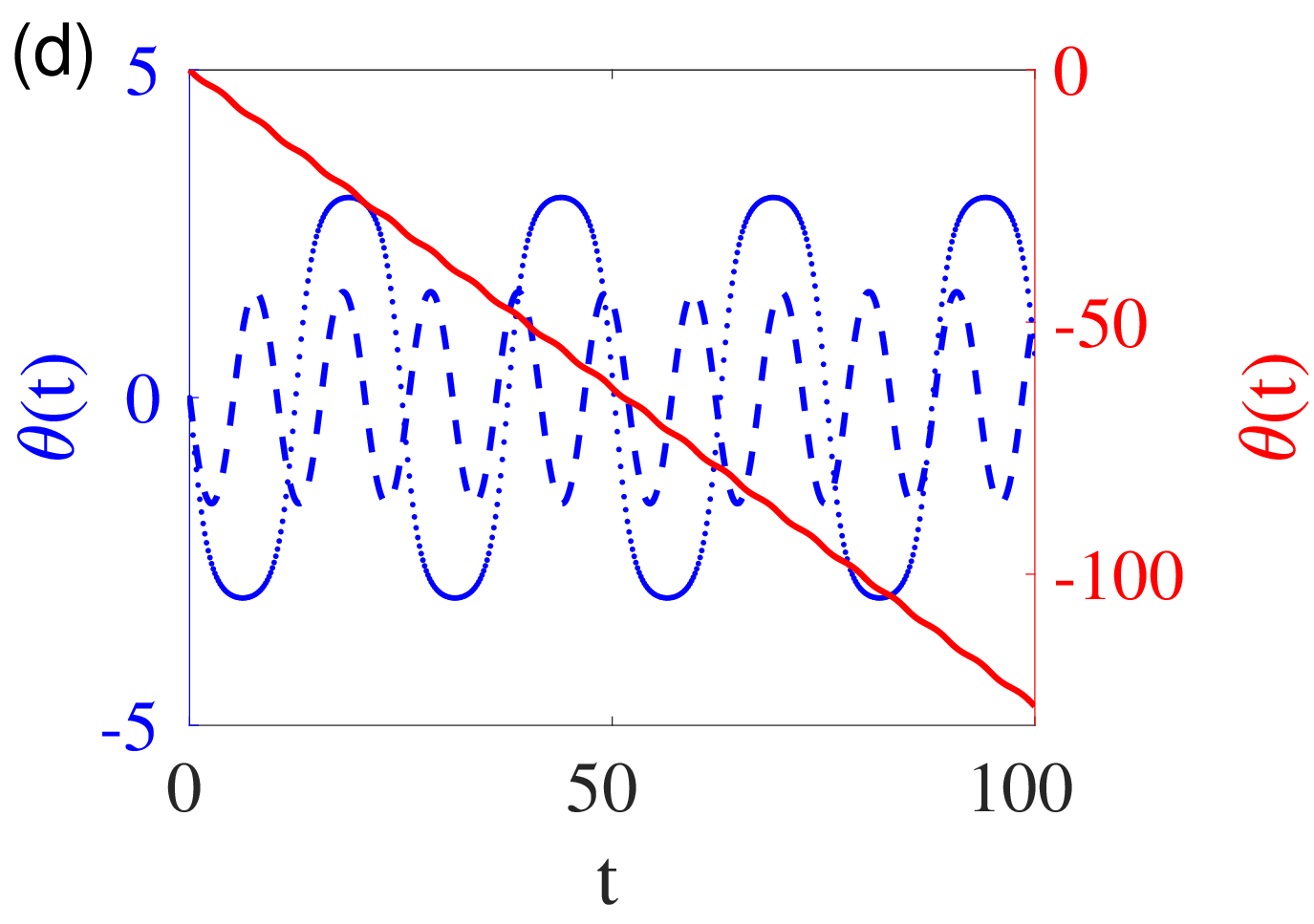}
\caption{The dynamics of the atomic imbalance $Z(t)$ for different initial values: $Z_0=0.04$ (a), $Z_{0c}=0.0553$ (b), and $Z_0=0.065$ (c). The solid and dashed lines correspond to the numerical simulations and the theoretical results, respectively. The phase dynamics corresponding to these initial values of the atomic imbalance are shown in panel (d). The dashed line ($Z_0=0.04$) and the dotted line ($Z_0=0.0553$) correspond to the left vertical axis, while the solid line ($Z_0=0.065$) corresponds to the right vertical axis. For all panels: $\theta_0=0$, $N=10$ $K=0.01$ and $g=1$.}
\label{fig2}
\end{figure}
Figure~\ref{fig2} (a), (b), and (c) show the time evolution of the atomic imbalance $Z(t)$ for the three initial imbalances $Z_0$ used in Figure~\ref{fig1} (a). Both numerical simulations (solid line) and theoretical predictions are presented (dotted line), demonstrating excellent agreement. The theoretical curves are obtained from the analytical solutions of Eqs.~(\ref{eq:dimerZtThetat}). A similar dynamical behaviour is observed in the $\pi$-phase mode. In Fig. \ref{fig2} (d), the phase dynamics corresponding to these values of the initial atomic imbalance are presented. The solid line represents the Josephson oscillations at $Z_0=0.04$, the dashed line indicates the critical trajectory at $Z_0=0.0553$, and the dotted line denotes the self-trapping mode at $Z_0=0.065$.

% Figure 3
\begin{figure}[htbp]
  \centerline{ \includegraphics[width=4.55cm]{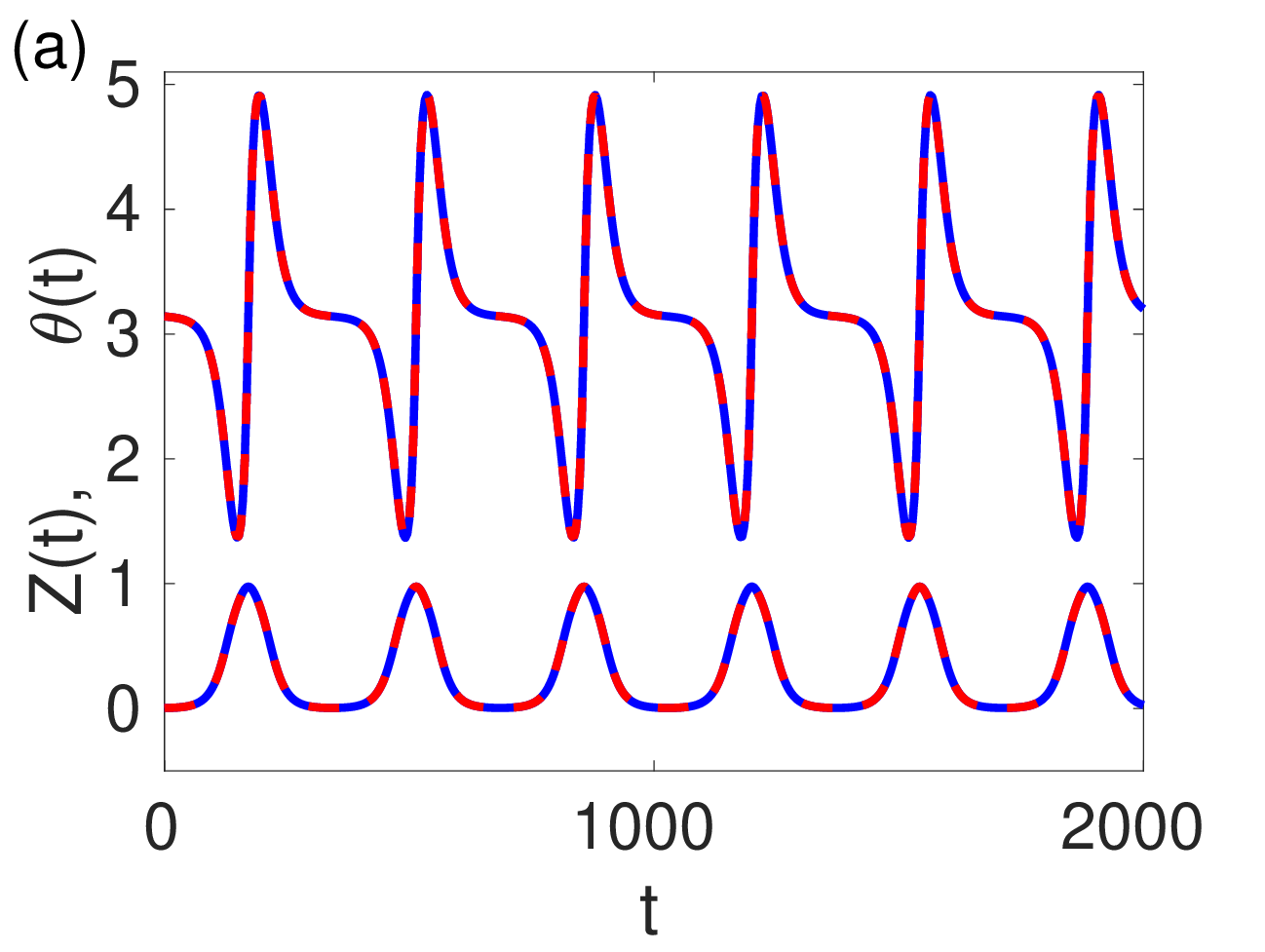}
  \includegraphics[width=4.55cm]{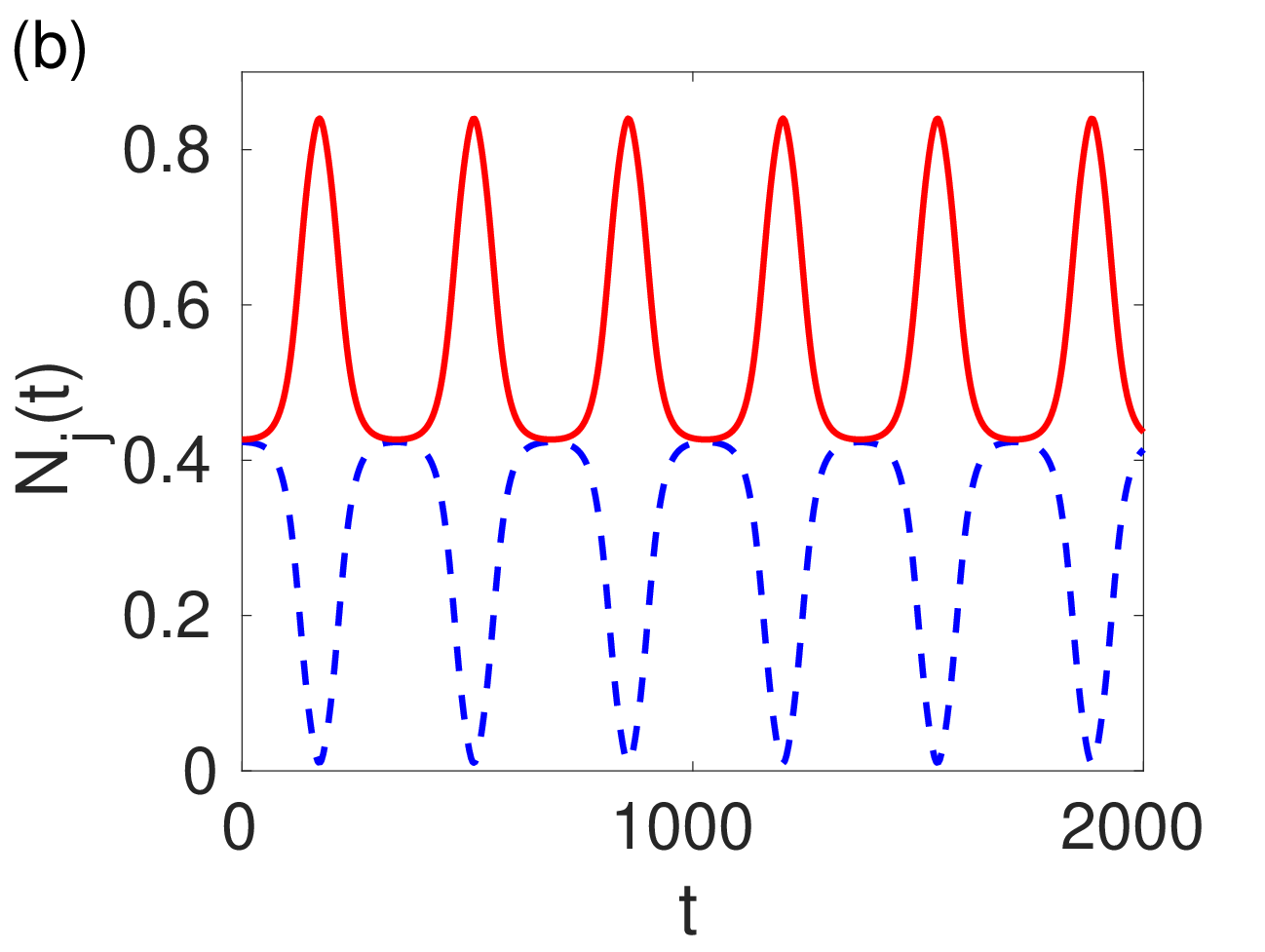}}
\caption{(a) The dynamics of $Z(t)$ (bottom plots) and relative phase $\theta(t)$ (upper plots) in the localisation revival mode are presented. The solid and dashed lines show the theoretical predictions and the numerical results, respectively. (b) Time evolution of the number of particles in each component. The bottom curve for $N_1$ and the upper curve for $N_2$. The parameters coincide with those of the dashed trajectory in the phase portrait of Fig.~\ref{fig1}(c).}
\label{fig-revival}
\end{figure}

Figure~\ref{fig-revival} (a) illustrates the dynamics of the atomic population imbalance and the relative phase. The population imbalance oscillates between $0$ and $1$, while the relative phase exhibits oscillations around its initial value $\theta_0$. This regime corresponds to the localisation revival regime. All particles transfer from the first component to the second, then return to the first until the populations equalise, and this cycle repeats periodically, see Fig.~\ref{fig-revival} (b). Unlike in the self-trapping mode, in the localisation revival regime, the value of the relative phase does not diverge to infinity. This nonlinear oscillation mode was observed in a one-dimensional ultradilute quantum liquid in a double-well potential \cite{Pol}, and we also observed it for a two-dimensional two-core BEC. 

% Figure 4
\begin{figure}[htbp]
  \centerline{ \includegraphics[width=4.55cm]{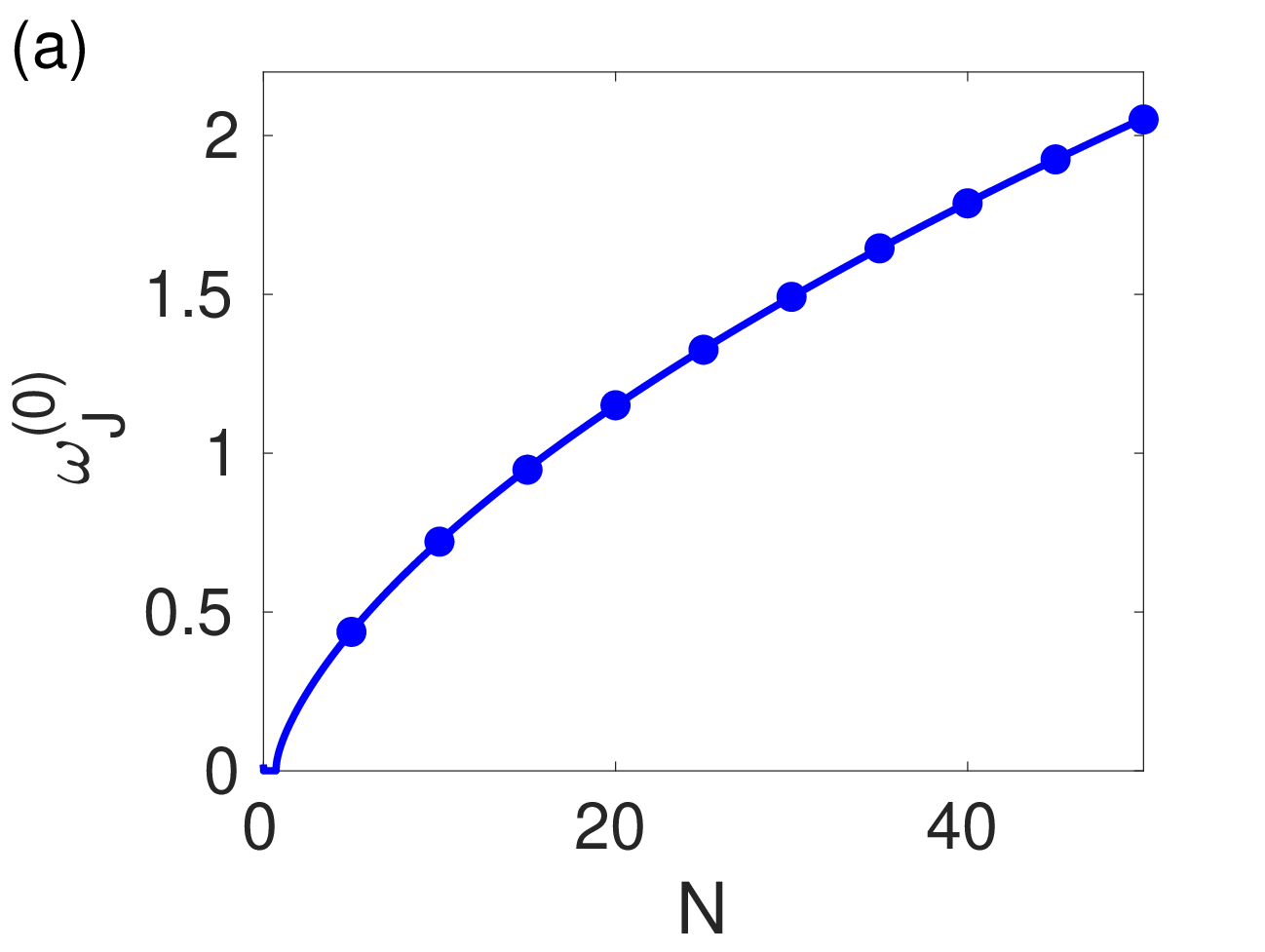}
  \includegraphics[width=4.55cm]{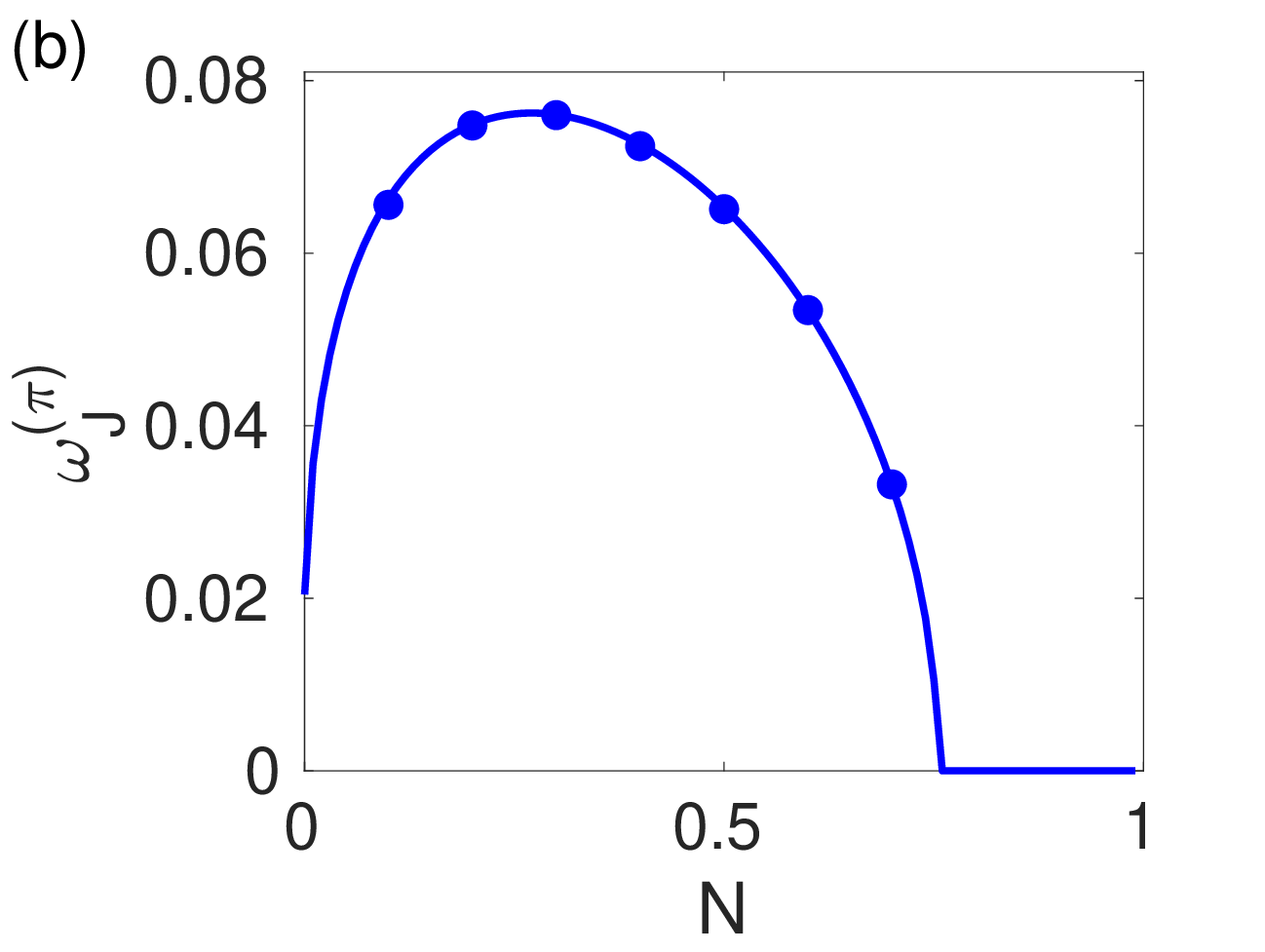}}
\caption{The dependence of the Josephson frequency on the number of atoms is presented for zero-(a) and $\pi$-phase (b) modes. The solid lines and the points correspond to the Eqs. (\ref{eq:Jfzero}, \ref{eq:Jfpi}) and numerical simulation results, respectively. Other parameters $K=0.01$ and $g=1$.}
\label{fig3}
\end{figure}

Figure \ref{fig3} shows a graph of the Josephson frequency versus the number of atoms based on Eqs. (\ref{eq:Jfzero}) and (\ref{eq:Jfpi}). It can be seen from the figures that in the zero-phase mode, Josephson oscillations exist, and as the number of atoms increases, the Josephson frequency also increases. However, for the $\pi$-phase regime, Josephson oscillations may exist in a narrow interval ($N<0.76$). When $N\geq0.76$, the Josephson frequency becomes complex, which corresponds to the self-trapping regime. In both panels, the VA and numerical results are in good agreement.

\section{Bifurcation}
\label{sec:bifurcation}

The fixed points $(Z^*, \theta^*)$ of Eq.~(\ref{eq:dimerZtThetat}) satisfy $Z^* = 0$ and $\theta^* = 0$. Setting $Z^* = 0$ yields $\theta^* = n\pi$, or alternatively $Z^* = \pm 1$. However, $Z^* = \pm 1$ leads to a singularity in $\theta_t$,  we discard this solution.

From the condition $\theta_t = 0$, we obtain
\begin{eqnarray}
&F(Z) = (1 - Z) \ln\left(\frac{N(1 - Z)}{2}\right) - \\
\nonumber 
& (1 + Z) \ln\left(\frac{N(1 + Z)}{2}\right)- \cfrac{2K Z \sigma}{\sqrt{1 - Z^{2}}},
\label{eq:FZ}
\end{eqnarray}
where \(\sigma = \cos(\theta^*) = \cos(n\pi) = (-1)^n = \pm 1\).

Evaluating the Jacobian matrix at the fixed point $(Z^*, \theta^*)$, we find
\begin{equation}
J^* =
\begin{pmatrix}
0 & 2K \sigma \sqrt{1 - Z^{*2}} \\
F_Z(Z^*)& 0
\end{pmatrix}.
\label{eq:Jacob}
\end{equation}

The trace and determinant of the Jacobian matrix are given by
$
\mathrm{Tr}(J^*) = 0,
$
\begin{equation}
\mathrm{det}(J^*) = -2K \sigma \sqrt{1 - Z^{*2}} F_Z(Z^*).
\label{eq:detJ}
\end{equation}

The sign of $\mathrm{det}(J^*)$ determines the type and stability of the fixed point. The eigenvalues are purely imaginary when $\mathrm{det}(J^*) > 0$; in this case, the fixed point is a center, which is neutrally stable, see Fig. \ref{fig1}(a). When $\mathrm{det}(J^*) < 0$, the eigenvalues are real and have opposite signs, indicating a saddle point, which is unstable, see Fig. \ref{fig1}(b). The case $\mathrm{det}(J^*) = 0$ corresponds to a bifurcation.

The bifurcation condition $\mathrm{det}(J^*) = 0$ leads to the critical particle number
\begin{equation}
N_{\mathrm{b}} = -\cfrac{2K \sigma}{g\, W_k \left(-\cfrac{eK \sigma}{g}\right)},
\label{eq:Ncr}
\end{equation}
here, $W_k(\cdot)$ denotes the Lambert $W$ function, where the subscripts $k=0$ and $k=-1$ refer to the principal and lower real branches, respectively. Since the real branches of the Lambert $W$ function exist only for arguments $\ge -\tfrac{1}{e}$, this requires
\begin{equation}
\frac{\sigma K}{g} \le e^{-2}.
\label{eq:Kg}
\end{equation}

\begin{figure}[htbp]
  \centerline{ \includegraphics[width=4.55cm]{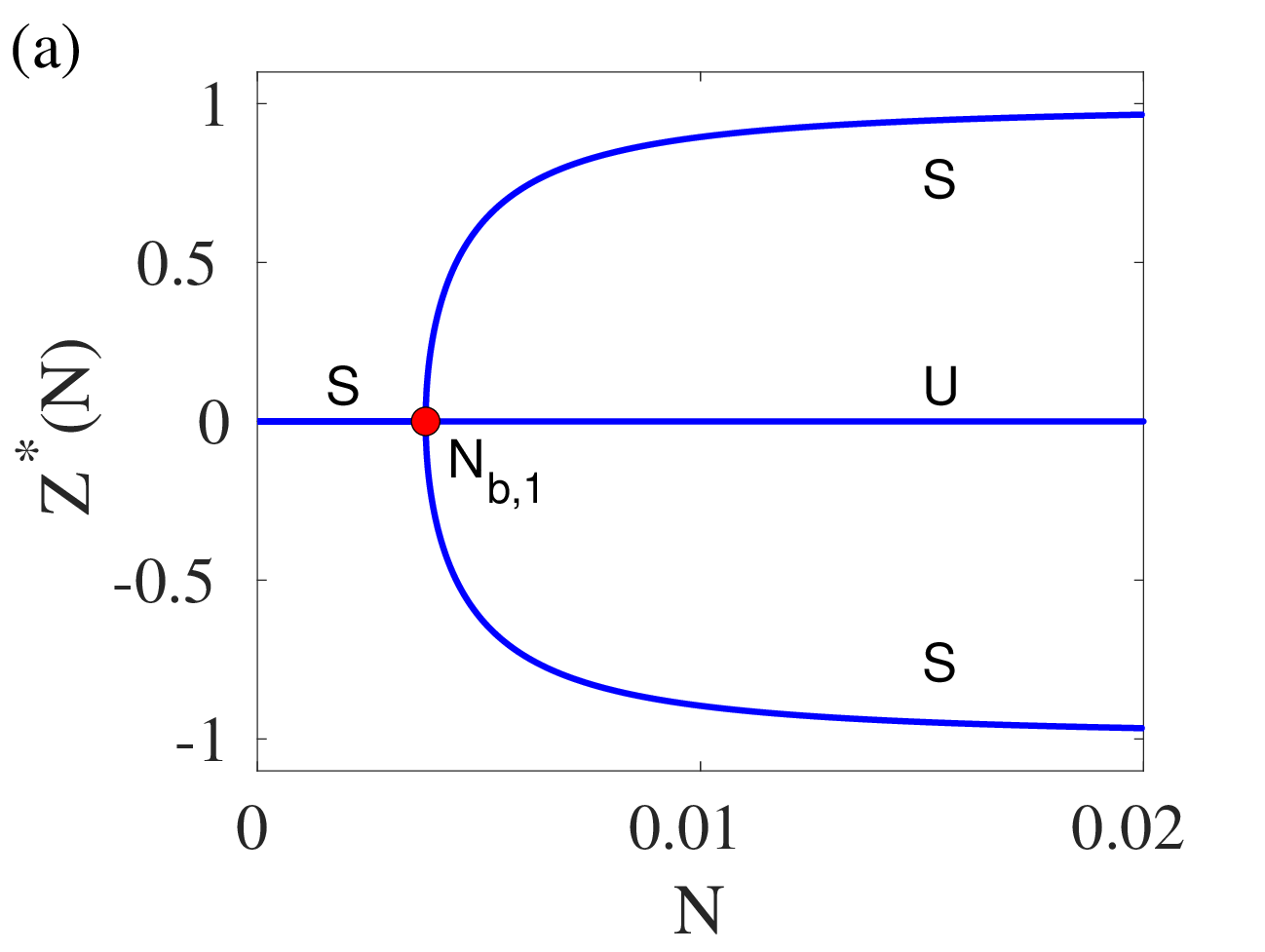}
  \includegraphics[width=4.55cm]{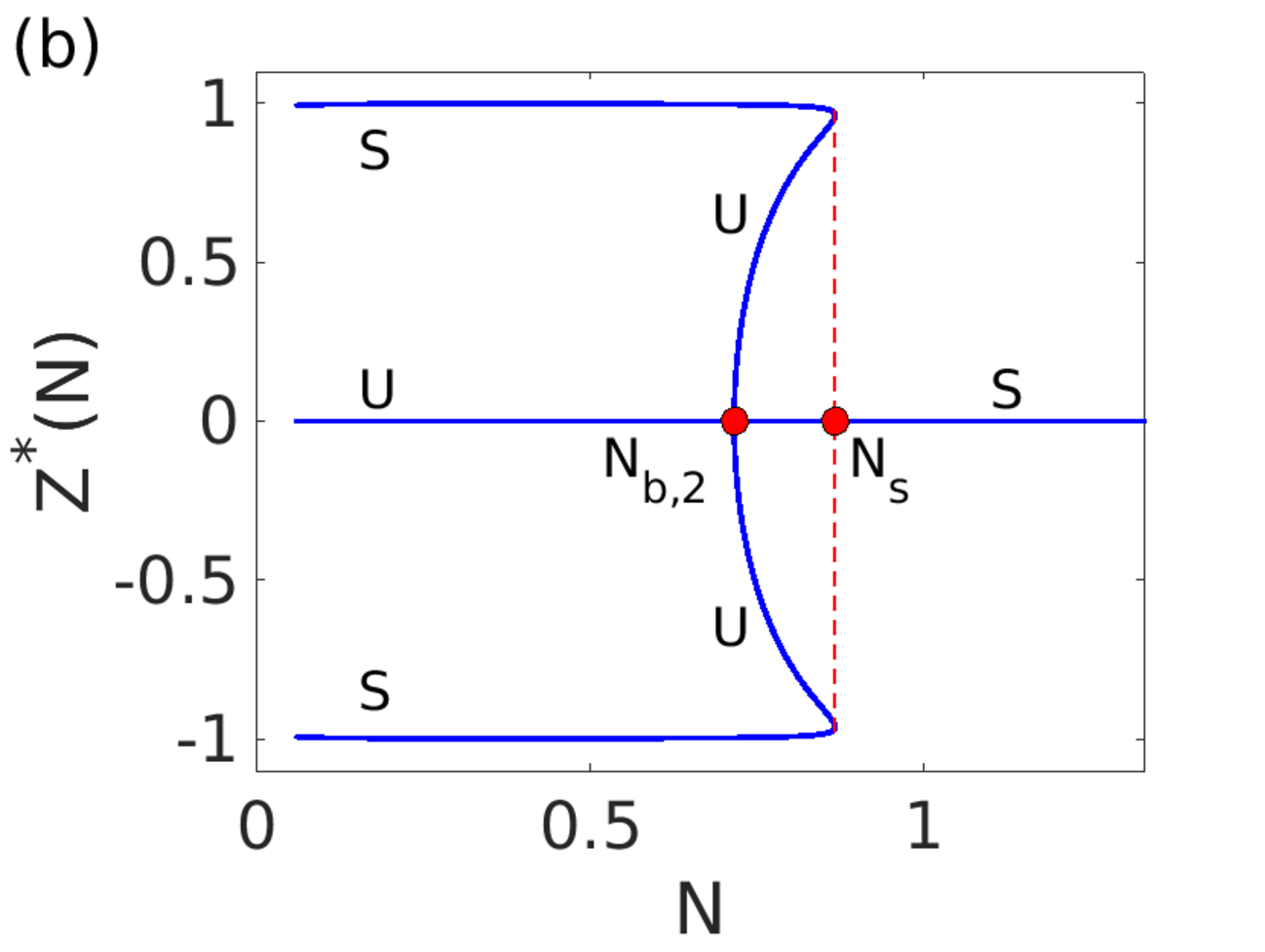}}
  \includegraphics[width=4.55cm]{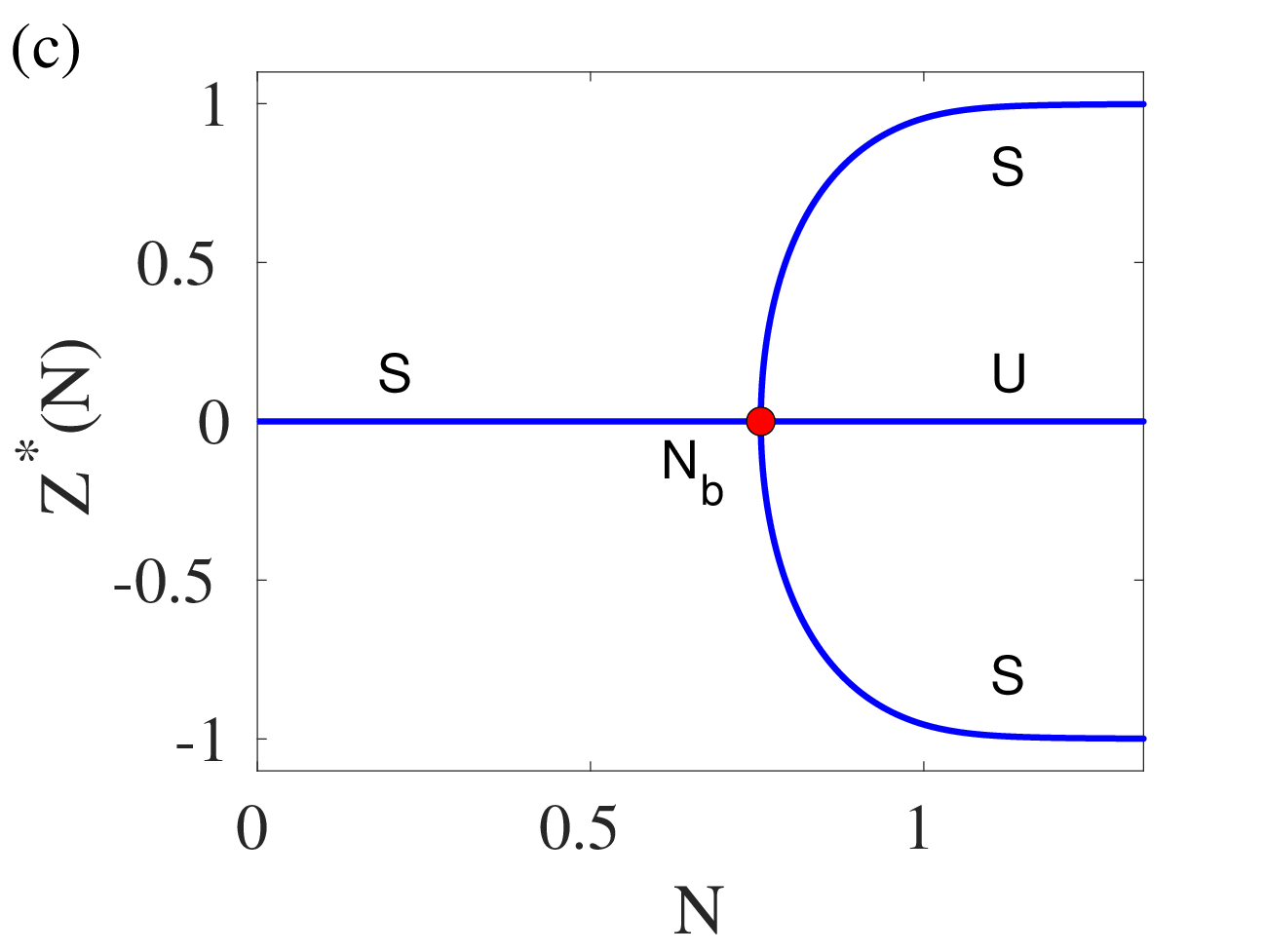} 
\caption{Bifurcation diagram in the $(N, Z)$-plane. 
(a) In the zero-phase, the first supercritical bifurcation occurs at $N_{b,1} \simeq 0.0038$, corresponding to $\sigma = 1$ in Eq.~(\ref{eq:Ncr}) and associated with the lower branch of the Lambert function $W_{-1}$. 
(b) As $N$ increases further in the zero-phase, the second subcritical bifurcation occurs at $N_{b,2} \simeq 0.7155$, determined by the principal branch $W_0$ of the Lambert function. A saddle-node bifurcation appears at $N = N_s \simeq 0.8665$, and hysteresis is observed in the interval $N_{b,2} < N < N_s$. 
(c) In the $\pi$-phase, a subcritical bifurcation occurs at $N_b \simeq 0.7555$, corresponding to $\sigma = -1$ and obtained via the principal branch $W_0$ in Eq.~(\ref{eq:Ncr}). The other parameters are fixed as $K = 0.01$ and $g = 1$.}
\label{fig:bifdiag}
\end{figure}

It follows from Eq.~(\ref{eq:Ncr}) that at $\sigma=1$ the zero-phase admits two real branches of the Lambert $W$ function, whereas the $\pi$-phase admits only one.  Consequently, there are two bifurcation points in the zero-phase and a single bifurcation point in the $\pi$-phase.  The corresponding bifurcation diagrams, obtained by numerically solving Eq.~(\ref{eq:dimerZtThetat}), are shown in Fig.~\ref{fig:bifdiag} (a) and (b) for the zero-phase, and in Fig.~\ref{fig:bifdiag} (c) for the $\pi$-phase.

% Figure 5

The figure~\ref{fig:bifdiag} clearly indicates that all bifurcations in the system are of the pitchfork type. Noticing that $F(-Z)=-F(Z)$ is odd, the equilibrium $Z=0$ exists for all $N>0$ and is the unique solution preserving the $Z\to -Z$ symmetry. In the following, we analyze the bifurcation diagrams for the zero-phase and $\pi$-phase separately.  

At the zero-phase, i.e., for $\sigma = 1$, there are two bifurcation points as discussed above. For the chosen parameters $g = 1$ and $K = 0.01$, the first bifurcation occurs at $N_{b,1} \simeq 0.0038$, which is marked in Fig.~\ref{fig:bifdiag}(a). This corresponds to a super-critical pitchfork bifurcation, in which the stable equilibrium at $Z^* = 0$ loses stability and gives rise to two stable symmetry-broken equilibria at $\pm Z^*$. In the bifurcation diagrams, stable and unstable branches are denoted by ``S'' and ``U'', respectively. 
Beyond this critical value, a second bifurcation of sub-critical type occurs. In this case, the bifurcation point corresponds to $N_{b,2} \simeq 0.7155$, as shown in Fig.~4(b). Here, the previously existing unbalanced equilibria $\pm Z^*$ (which are now unstable) collide with $Z^* = 0$ and disappear, thereby restoring the stability of the origin.
In the interval $N_{b,2} < N < N_s \simeq 0.8665$, different distinct stable equilibria coexist: the symmetric point $Z^* = 0$ and a pair of large-imbalance fixed points. Consequently, the long-term state reached as $t \to \infty$ depends on the initial condition $Z_0$; whichever basin of attraction contains $Z_0$ determines the outcome. This multistability permits discontinuous jumps and a hysteresis loop when the control parameter $N$ is varied slowly.
Let us start the system in the symmetric state $Z^{*} = 0$ at some $N > N_s$ and decrease $N$ quasi‑statically. The trajectory remains at the origin until $N$ reaches $N_{b,2}$, at which point the trivial equilibrium loses stability. An infinitesimal perturbation then sends the system to one of the symmetry‑broken, large‑imbalance branches. Further reduction of $N$ simply carries the state along this branch. 
If $N$ is now increased again, the solution persists on the large‑imbalance branch even after $N$ rises past $N_{b,2}$. Only when $N$ reaches $N_s$ does the large‑imbalance branch terminate in a saddle‑node bifurcation, forcing the system to jump back to the origin. This path‑dependence, where the response to a cyclic variation of $N$ is irreversible, is precisely what is known as hysteresis~\cite{Strogatz2018}, see Fig.~\ref{fig:bifdiag} (b).

Figure~\ref{fig:bifdiag} (c) depicts the bifurcation diagram in the $\pi$‑phase, which corresponds to the subcritical pitchfork bifurcation. Here, the bifurcation occurs at $N_{b} \simeq 0.7555$, corresponding to the principal branch of the Lambert $W$ function for $\sigma = -1$ in Eq.~(\ref{eq:Ncr}).

\begin{figure}[htbp]
  \centerline{ \includegraphics[width=4.55cm]{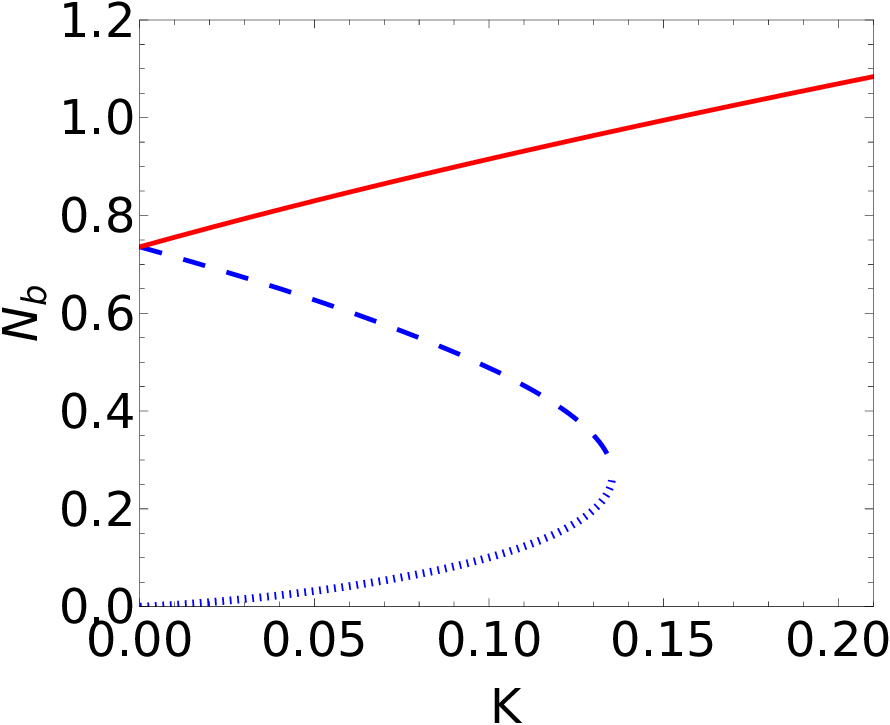}}
\caption{Dependence of the bifurcation points $N_b$ on the coupling constant $K$ for fixed $g = 1$. 
The red solid line at the top corresponds to the $\pi$-phase mode ($\sigma = -1$). 
The blue dashed line and blue dotted markers represent the principal and lower branches of the Lambert function $W$ in the zero-phase mode ($\sigma = 1$), respectively; see Eq.~(\ref{eq:Ncr}).}
\label{fig:NbvsK}
\end{figure}

Figure~\ref{fig:NbvsK} plots the bifurcation points (in terms of $N$) against the coupling constant $K$, see Eq.~(\ref{eq:Ncr}). The solid red curve denotes the $\pi$-phase; the dashed blue curve represents the principal branch of the Lambert $W$ function in the zero-phase; and the dotted blue markers indicate its lower branch. 

In the zero-phase, bifurcations occur only within finite regions of the $(N, K)$-plane, with the turning point at  
$K = 0.135335$, see Eq.~(\ref{eq:Kg}). By contrast, in the $\pi$-phase, bifurcations persist for all $K$. 
One can analyse the bifurcations in the system by its normal form. To obtain the normal form, we expand Eq.~(\ref{eq:dimerZtThetat}) in a Taylor series about $Z^*=0$. Carrying the expansion out to the fifth order, one finds
\begin{equation*}
F(Z)\simeq \alpha(N) Z +\beta(N) Z^3 +\gamma(N) Z^5+\mathcal{O}(Z^7),
\end{equation*}
with
\begin{align*}
\nonumber
\alpha(N) &= -2K\sigma - g \, N\, \ln\!\bigl(eN/2\bigr),\\
\nonumber
\beta(N)  &= \tfrac{g\,N}{6} - K\sigma,\\
\gamma(N) &= \tfrac{g\,N}{20} - \tfrac{3K}{4}.
\end{align*}
Retaining only the leading-order terms, $F(Z) \approx \alpha(N) Z +\beta(N) Z^3$, recovers the canonical normal form of a pitchfork bifurcation~\cite{Strogatz2018}.
In the following section, we examine Josephson oscillations between quantum droplets.

\section{Droplets interaction}
\label{sec:QDinteraction}

We employ the variational approximation (VA) to investigate macroscopic quantum tunnelling between quantum droplets. The corresponding Lagrangian density for the system described by Eq.~(\ref{eq:coupGPE}) is given by:
\begin{eqnarray}\label{eq:lagr}
&\mathcal{L} = \frac{i}{2}(uu_t^{\ast} - u^{\ast}u_t) +\frac{1}{2}|\nabla u|^2 +\frac{g}{2}|u|^4\ln(|u|^2) - \nonumber\\
&\frac{g|u|^4}{4}-K(u^{\ast}v + u v^{\ast}) + \frac{i}{2}(vv_t^{\ast} - v^{\ast}v_t)+ \nonumber\\
&\frac{1}{2}|\nabla v|^2 +\frac{g}{2}|v|^4\ln(|v|^2)-\frac{g|v|^4}{4}.
\end{eqnarray}
Following the idea of the work~\cite{PF}, let us investigate the Josephson oscillations and self-trapping regimes, using the VA to the QD dynamics.
The VA  ansatz for the wavefunctions is~\cite{Otajonov2019Lavoine2021, Baizakov2011Otajonov2024, Otajonov2020}:
\begin{eqnarray}\label{eq:ansatz}
u=A_1 \exp\left(-\frac{1}{2} (ar)^{2m} + i\phi_1 \right), \nonumber\\
v = A_2 \exp\left(-\frac{1}{2}(ar)^{2m} +i\phi_2\right).
\end{eqnarray}
The averaged Lagrangian is given by $L = \int_{-\infty}^{\infty} \mathcal{L} dxdy$.
Substituting the ansatz (\ref{eq:ansatz}) into Eq.(\ref{eq:lagr}) and integrating, we found the averaged Lagrangian:

\begin{eqnarray}
L &=& \frac{\pi }{a^2} \left\{ \sum_{i=1}^{2}\left[\Gamma(1+M)A_i^2\phi_{i,t}^2 
+ \frac{1}{2M}a^2A_i^2 \right.\right. \nonumber \\
&& \left.\left. - 2^{-2-M}g \Gamma(1+M)((1+M)A_i^4 - 2A_i^4\ln A_i^4)\right] \right. \nonumber\\
&& \left. -2K A_1 A_2\cos(\phi_2-\phi_1) \Gamma(1+M) \right\},
\end{eqnarray}

where $M=1/m$.
The Euler-Lagrange equations give:
\begin{eqnarray} \label{eq:QD_JosepEq}
&\theta_t=-\frac{2KZ}{\sqrt{1-Z^2}}\cos{\theta}+\frac{2^{-1-M}a^2gN}{\pi\Gamma(1+M)} -\left[ZM \nonumber \right.\\
&(1+Z)\ln{\left(\frac{N(1+Z)}{2}\right)}+(1-Z)\ln{\left(\frac{N(1-Z)}{2}\right)} \nonumber\\
&\left.-2Z\ln{\left(\frac{a^2}{\pi \Gamma(1+M)}\right)}\right] \nonumber \\
&Z_t=2K\sqrt{1-Z^2}\sin{\theta},
\end{eqnarray}
where $\theta=\phi_2-\phi_1$ is relative phase, $Z=(N_2-N_1)/N$ is the atomic imbalance, $N_1=2\pi\int_{0}^{\infty}|u|^2dr=2^{-2-M}\pi g A_1^4 \Gamma(1+M)/a^2$, $N_2=2\pi\int_{0}^{\infty}|v|^2dr = 2^{-2-M}\pi g A_2^4 \Gamma(1+M)/a^2$ are the number of atoms in the cores and $N$ is the total number of atoms. We can write the Hamiltonian function for these new variables as:
\begin{eqnarray} \label{Hamiltonian}
   & H=2K\sqrt{1-Z^2}\cos{\theta}+\frac{2^{-2-M}a^2gNZ^2}{\pi \Gamma(M)}\left[1+\frac{1}{M}- \right. \nonumber\\
   & \left.\frac{2}{M} \ln{\left(\frac{a^2}{\pi \Gamma(1+M)}\right)}\right]-\frac{2^{-2-M}a^2gN}{\pi \Gamma(1+M)}\left[(1-Z)^2\times \right. \nonumber\\
   &\left. \ln{\left(\frac{N(1-Z)}{2}\right)}+(1+Z)^2\ln{\left(\frac{N(1+Z)}{2}\right)} \right].
\end{eqnarray}
Using the system of equations (\ref{eq:QD_JosepEq}), it is possible to find analytical expressions for Josephson frequencies for small amplitude oscillations. a) zero-phase mode:
\begin{eqnarray} \label{JFzero}
    \omega_J=2K \left\{1-\frac{2^{-M}a^2gN}{2K\pi M \Gamma(M)}\times \right. \nonumber\\ 
    \left.\left[\frac{M}{2}-\ln{\left(\frac{e a^2N}{2\pi M \Gamma(M)}\right)}\right]\right \}^{1/2},
\end{eqnarray}
b) $\pi$ -phase mode:
\begin{eqnarray}  \label{JFpi}
    \omega_J=2K \left\{1+\frac{2^{-M}a^2gN}{2K\pi M \Gamma(M)}\times \right. \nonumber\\ 
    \left.\left[\frac{M}{2}-\ln{\left(\frac{e a^2N}{2\pi M \Gamma(M)}\right)}\right]\right \}^{1/2}.
\end{eqnarray}

We can derive the self-trapping condition using Eq.~(\ref{Hamiltonian}):
\begin{eqnarray}
    H(Z_0,\theta_0)>\pm 2K-\frac{2^{-1-M}a^2gN\ln{(N/2)}}{\pi \Gamma(1+M)},
\end{eqnarray}
and can be found $K_{\mathrm{cr}}$:
\begin{eqnarray} \label{critical_v}
    &K_{\mathrm{cr}}=\frac{2^{-3-M}Ma^2gN\ln{(N/2)}}{\pi \Gamma(1+M)(\pm 1-\sqrt{1-Z_0^2}\cos{\theta_0}}\left\{\frac{2}{M}\ln{\left(\frac{N}{2}\right)}+\right. \nonumber\\
   & Z_0^2-\frac{2Z_0^2}{M}\ln{\left(\frac{a^2}{\sqrt{e}\pi \Gamma(1+M)}\right)}-\frac{1}{M} \left[(1-Z_0)^2 \right. \nonumber \\
   & \ln{\left(\frac{N(1-Z_0)}{2}\right)}+(1+Z_0) \ln{\left(\frac{N(1+Z_0)}{2}\right)} \Bigr] \Bigr \},
\end{eqnarray}
here, the $\pm$ signs correspond to the zero- and $\pi$ -phase modes, respectively. In the symmetric case, Eq.~(\ref{eq:coupGPE}) can be rewritten as
\begin{equation*}
i\,\partial_{t} u + \frac{1}{2} \nabla^{2} u 
- g\,|u|^{2} \log\left(|u|^{2}\right) u + K u = 0.
\end{equation*}
The linear coupling term can be eliminated by applying the transformation $u(\mathbf{r},t) = \Psi(\mathbf{r},t) \exp(-i K t)$. Here, the new wave function $\Psi$ inherits exactly the same modulus, i.e., $|\Psi| = |u|$, so the nonlinear term remains unchanged.
The stationary and dynamical properties of the resulting equation, 
analysed within the VA using a super-Gaussian ansatz, have been thoroughly studied in Ref.~\cite{Otajonov2020}. Therefore, in our subsequent theoretical analysis in this section, as well as in solving Eq.~(\ref{eq:QD_JosepEq}), we implement the variational results obtained in Ref.~\cite{Otajonov2020} to determine the stationary parameters of quantum droplets. It is worth noting that, in the symmetric case, the VA provides highly accurate predictions for the stationary parameters of quantum droplets, in very good agreement with numerical simulations.

\begin{figure}[htbp]
  \centerline{ \includegraphics[width=4.55cm]{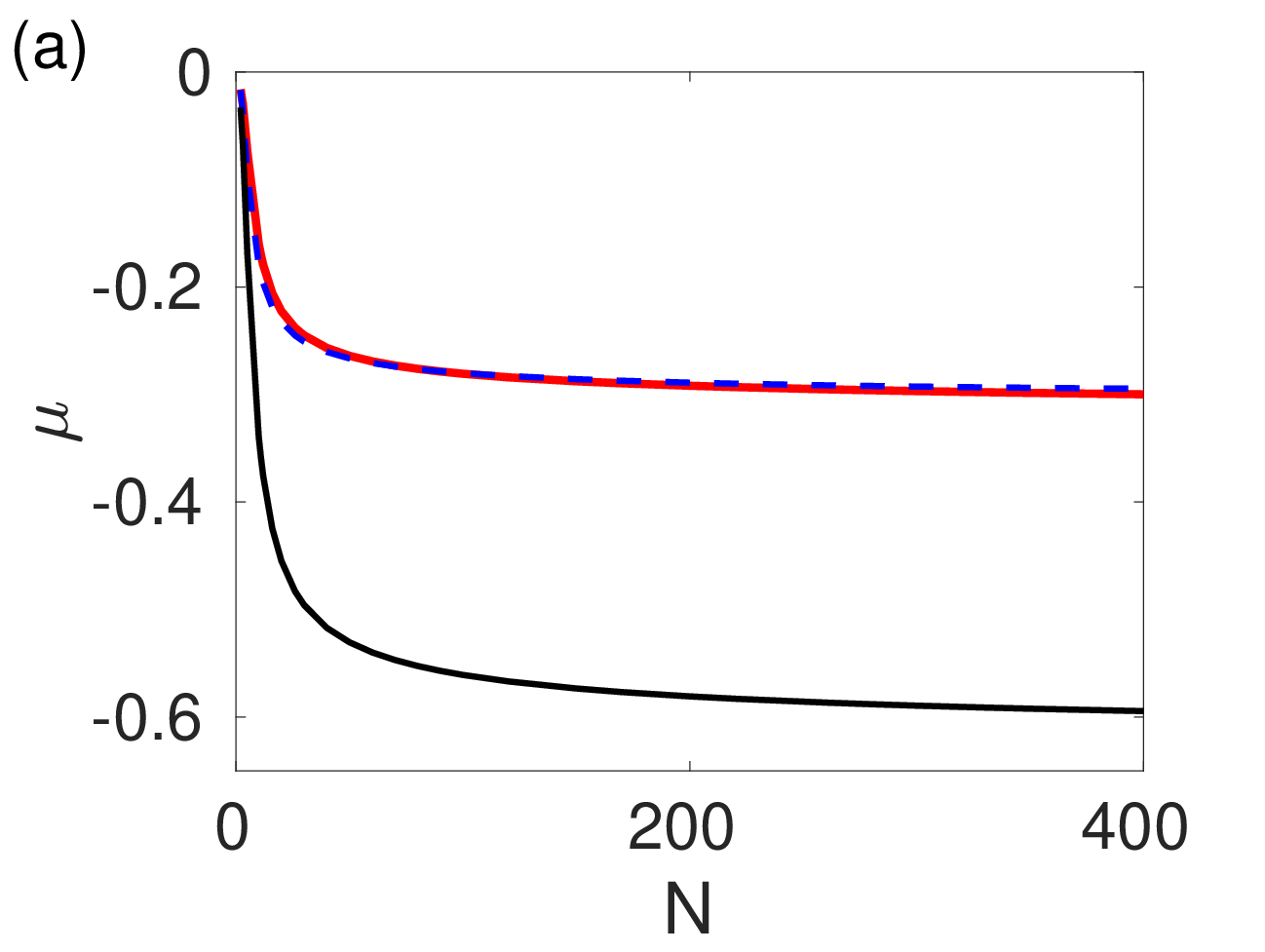}
  \includegraphics[width=4.4cm]{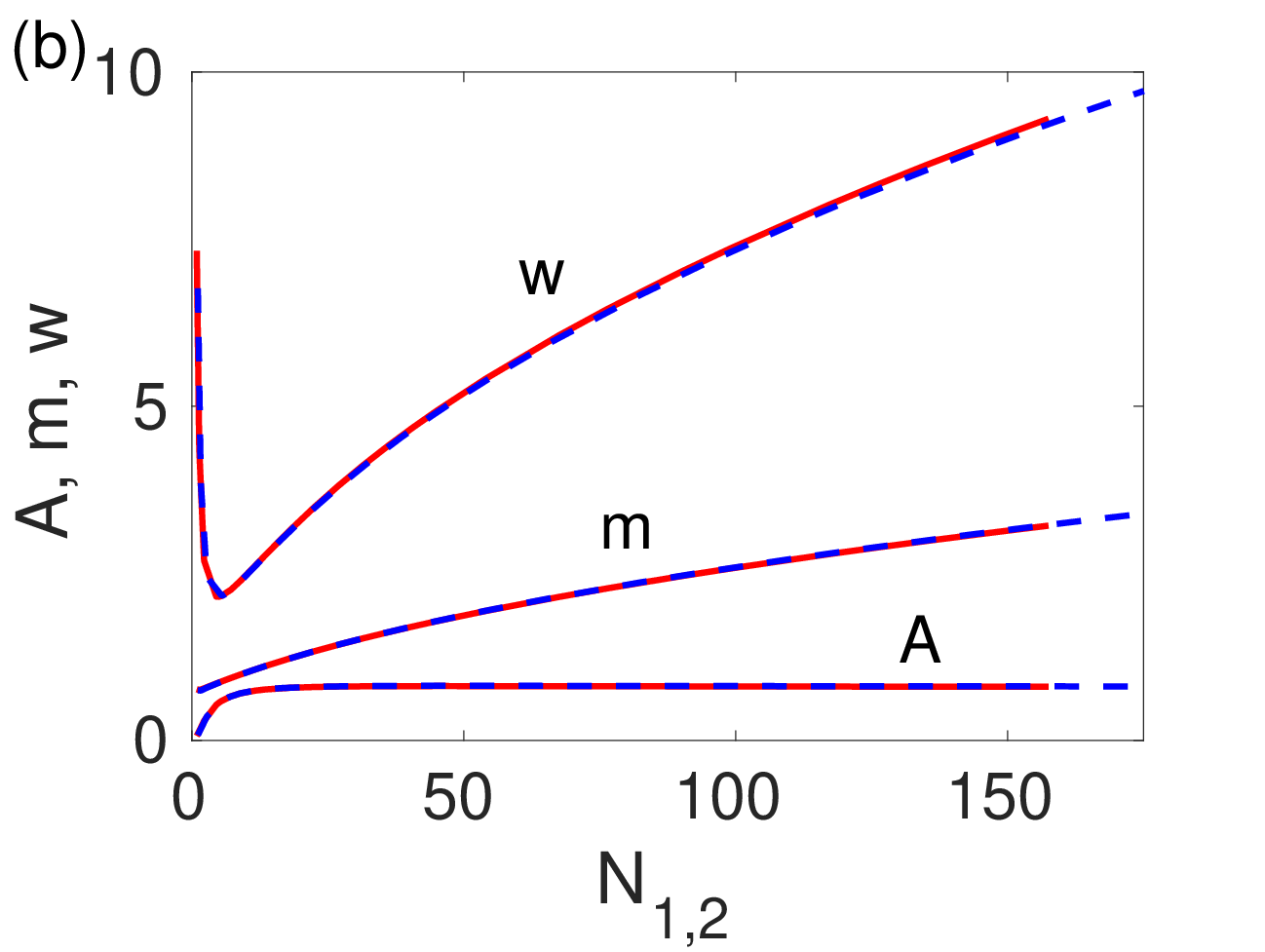}}
\caption{(a) Dependence of the chemical potential $\mu$ on the particle number $N$. The red solid and blue dashed curves correspond to the chemical potentials of the first $\mu_{1}(N_{1})$ and second $\mu_{2}(N_{2})$ components, respectively, while the bottom black solid curve represents the total chemical potential $\mu(N)$. (b) Stationary asymmetric QD parameters (amplitude, width, and super-Gaussian indices) as functions of $N$. The red solid and blue dashed curves correspond to the parameters of the first and second components, respectively. 
In both panels, the results of imaginary-time simulations of Eq.~(\ref{eq:coupGPE}) are presented with initial population imbalance $Z_{0}=0.1$. Other parameters are $K=0.01$ and $g=1$.}
\label{fig:muNParam}
\end{figure}

Figure~\ref{fig:muNParam} (a) shows the dependence of the chemical potential on the number of atoms $N$, obtained from imaginary-time simulations of Eq.~(\ref{eq:coupGPE}) with an initial imbalance $Z_0 = 0.1$ between the components. The red solid line represents the chemical potential $\mu_1(N_1)$ of the first component $u_1$, while the blue dashed line represents $\mu_2(N_2)$ of the second component $u_2$. In the numerical computations, the individual component norms are defined as $N_1 = N(1 - Z_0)/2$ and $N_2 = N(1 + Z_0)/2$, respectively. The bottom black solid line illustrates the total chemical potential $\mu = \mu_1 + \mu_2$ as a function of the total norm $N = N_1 + N_2$. According to the Vakhitov-Kolokolov stability criterion, the negative sign of the derivative $d\mu/dN < 0$ corresponds to dynamically stable quantum droplets. This stability was further confirmed by direct numerical simulations under small perturbations.

% Figure 7

\begin{figure}[htbp]
  \centerline{ \includegraphics[width=7.55cm]{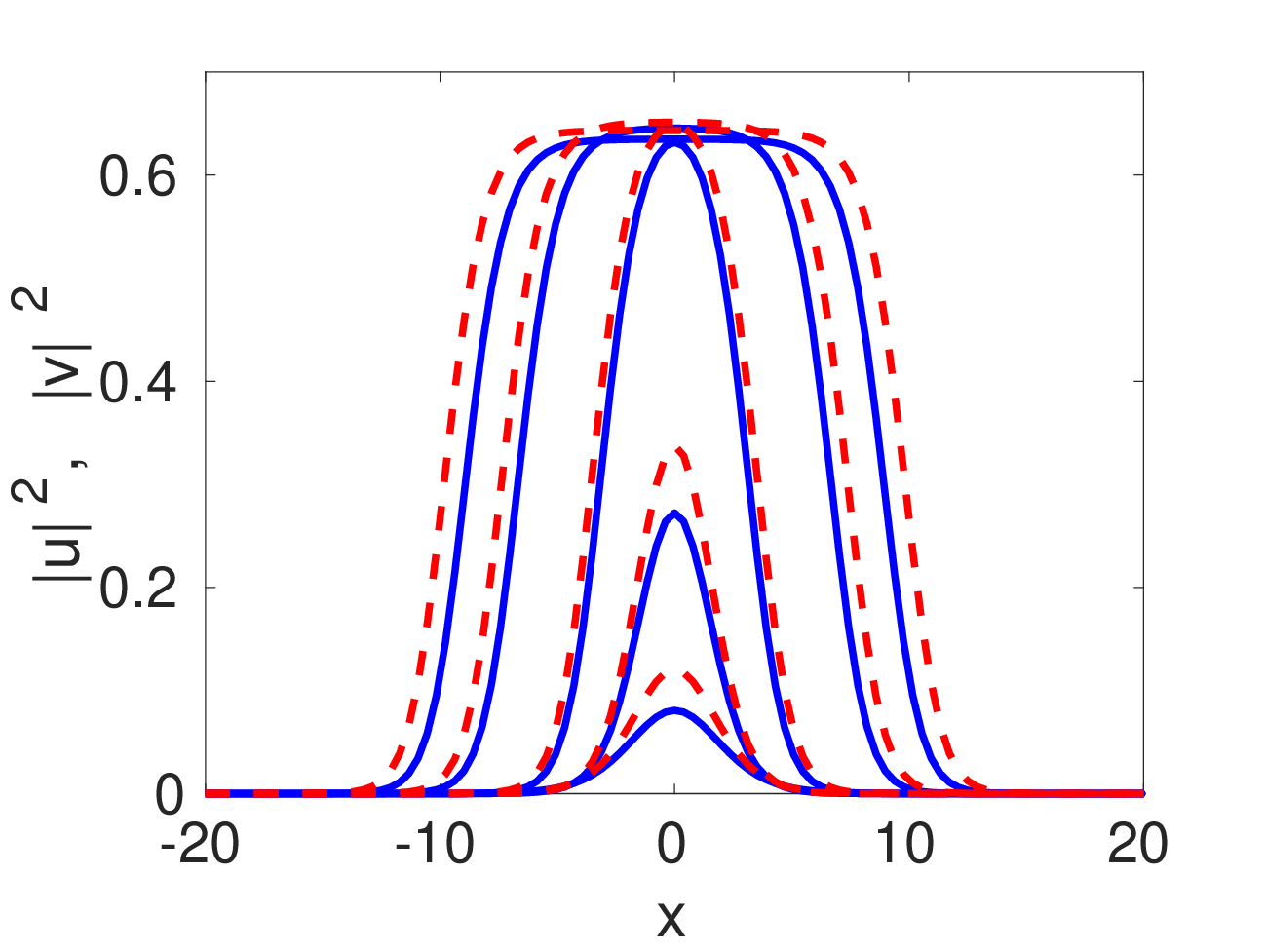}}
\caption{Profiles of the two components for different values of $N$. 
The blue solid curves represent the QD profiles of the first component $|u|^2$, 
while the red dashed curves correspond to those of the second component $|v|^2$. 
From bottom to top, the curves correspond to $N = 5, 10, 50, 200,$ and $350$. 
The other parameters are fixed as $Z_0 = 0.1$, $K = 0.01$, and $g = 1$.}
\label{fig:denUV} 
\end{figure}

Figure~\ref{fig:muNParam} (b) presents the dependence of parameters of stationary quantum droplets, namely, amplitudes, widths, and super-Gaussian indices, on the number of particles in each component, for an initial imbalance $Z_0 = 0.1$. The red solid line and the blue dashed line correspond to the first and second components, respectively. As the number of particles increases, the amplitudes of both components saturate, while their widths begin to grow noticeably only at higher norms. The increase in the super-Gaussian index $m$ with particle number indicates a transition of the droplet profile from a Gaussian to a flat-top shape. This behaviour highlights the emergence of an incompressible fluid-like nature in the quantum droplets (see also Fig.~\ref{fig:denUV}), consistent with the findings of Ref.~\cite{PA}.

% Figure 8

Figure~\ref{fig:denUV} displays the density profiles of both components for various values of total norm $N$, assuming $Z_0 = 0.1$. For small $N$, the droplet profiles are approximately Gaussian, whereas for larger $N$, the profiles develop flat-top structures, further supporting the identification of the droplets as incompressible quantum fluids.

% Figure 9
\begin{figure}[htbp]
  \centerline{ \includegraphics[width=4.55cm]{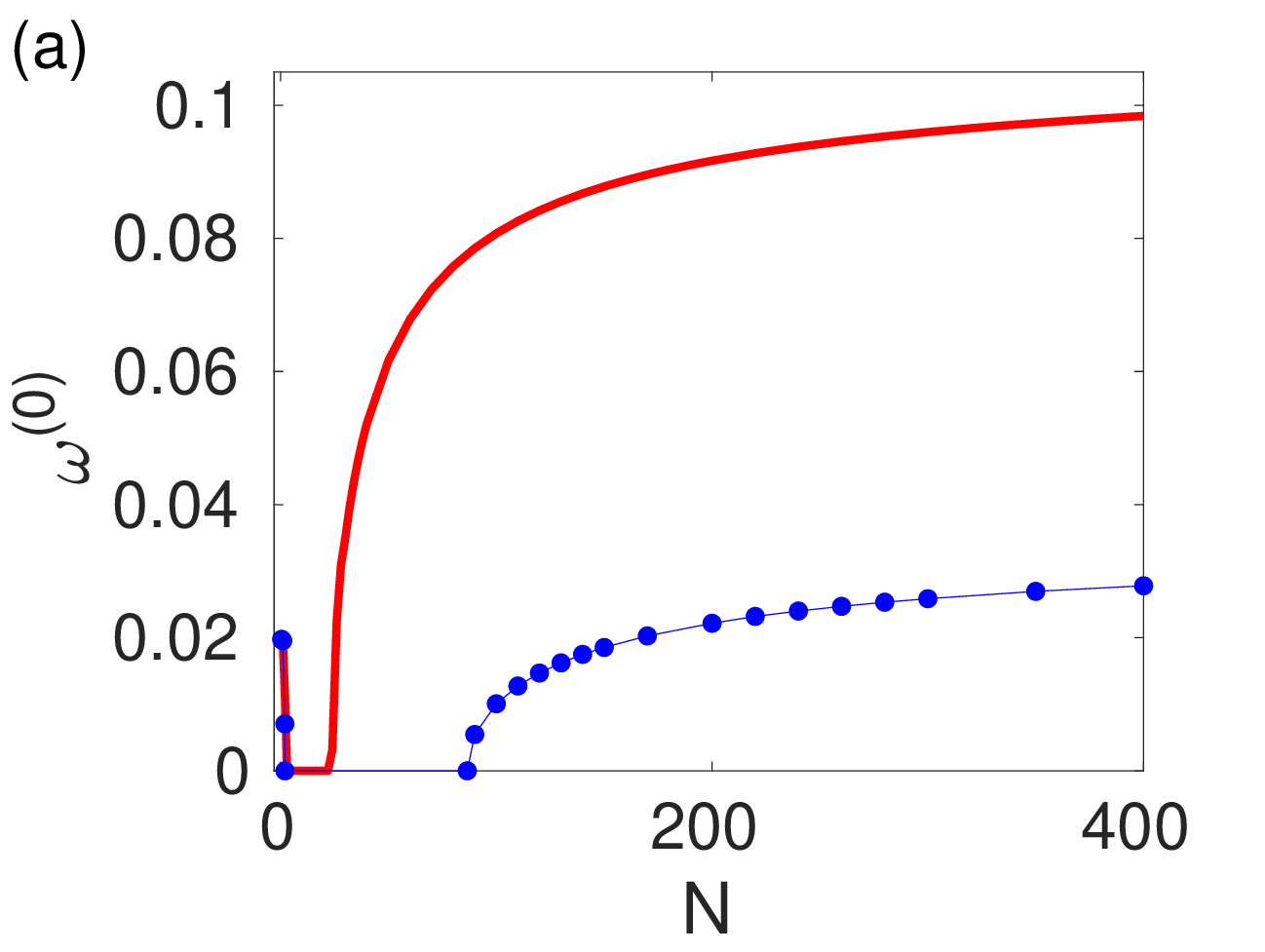}
  \includegraphics[width=4.4cm]{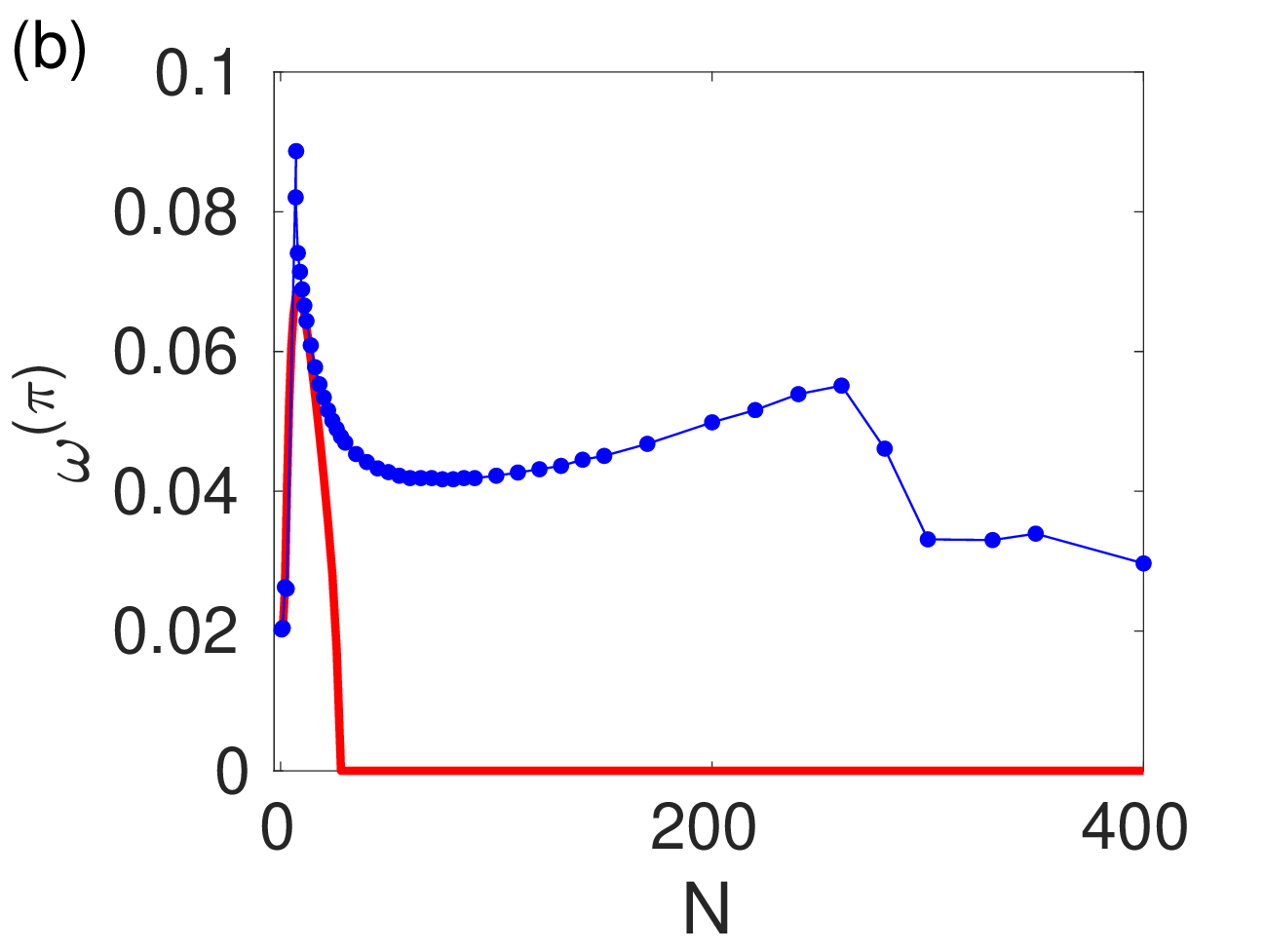} }
\caption{The dependence of the Josephson frequencies $\omega$ on the norm $N$ is presented. Graphs are shown for both the zero-phase (a) and $\pi$-phase (b) regimes. The solid lines represent the theoretical predictions based on Eqs. (\ref{JFzero}) and (\ref{JFpi}), while the dots correspond to numerical results. In both panels, the initial population imbalance is set to $Z_0 = 0.1$, with parameters $K = 0.01$ and $g = 1$.}
\label{fig-freq}
\end{figure}

Figure \ref{fig-freq} (a) displays the Josephson frequency as a function of the norm for the zero-phase regime. The dashed curve represents the theoretical results obtained from Eq. (\ref{JFzero}), while the points joined by a line represent the numerical simulation results. As shown in the figure, the theoretical and numerical results agree well for a small number of atoms. As the number of atoms increases, the overall trend remains similar, but the actual values begin to differ significantly. At certain values of the number of atoms, the Josephson frequency obtained from Eq. (\ref{JFzero}) becomes a complex number, leading to a discontinuity in the frequency curve. In theory, the part of the Josephson frequency where it is a complex number is represented as zero on the graph for these specific frequency values. According to theoretical results, this gap appears within the norm range of approximately $2.25$ to $23.70$, while numerical simulations indicate a broader interval from $2.25$ to $86.54$. This range of numbers of atoms corresponds to the self-trapping regime. However, this interval was derived for fixed values of the parameters $g$ and $K$. As can be seen from Eq. (\ref{JFzero}), within this interval of $N$, where self-trapping occurs, the Josephson frequency becomes complex. Nevertheless, by appropriately tuning the parameters $K$ and $g$, it is still possible to observe Josephson oscillations in this interval of $N$. A similar behaviour is found in the interval obtained from the numerical simulations, where Josephson oscillations are also present, provided that the parameters $K$ and $g$ are chosen appropriately.

Figure \ref{fig-freq} (b) corresponds to the $\pi$-phase regime. In this case, the results obtained from the VA agree well with the numerical results only when the number of atoms is sufficiently small ($N<16$). Numerical simulations reveal that a self-trapping regime occurs within the range of the number of atoms from $2.85$ to $7.22$. When $N>7.22$, quantum droplets, after a few oscillation periods up to $5-7$, move away from each other. This separation is attributed to the interaction energy becoming repulsive, which occurs because it is proportional to $\cos{\theta}$ (see Eq. (\ref{eq:int})), where $\theta$ is the relative phase of QDs.

For large particle numbers $N$, the results of the VA and numerical simulations show noticeable discrepancies. The primary reason is that, in deriving the Josephson equations, we assumed identical super-Gaussian parameters for both quantum droplets, i.e., $m_1=m_2=m$, as well as equal widths, $w_1=w_2=1/a$, which were considered time-independent, while only the amplitudes were allowed to vary in time. In reality, as the number of atoms increases, the super-Gaussian index $m$ and the droplet width $w=1/a$ both undergo significant changes, whereas the amplitude remains almost unaffected. Consequently, the discrepancy between the VA and numerical results becomes more pronounced as $N$ increases. If one were to incorporate the time dependence of the super-Gaussian parameters and the droplet width into the VA, the derivation of the Josephson equations would become mathematically intractable.

% Figure 10
\begin{figure}[htbp]
  \centerline{ \includegraphics[width=4.5cm]{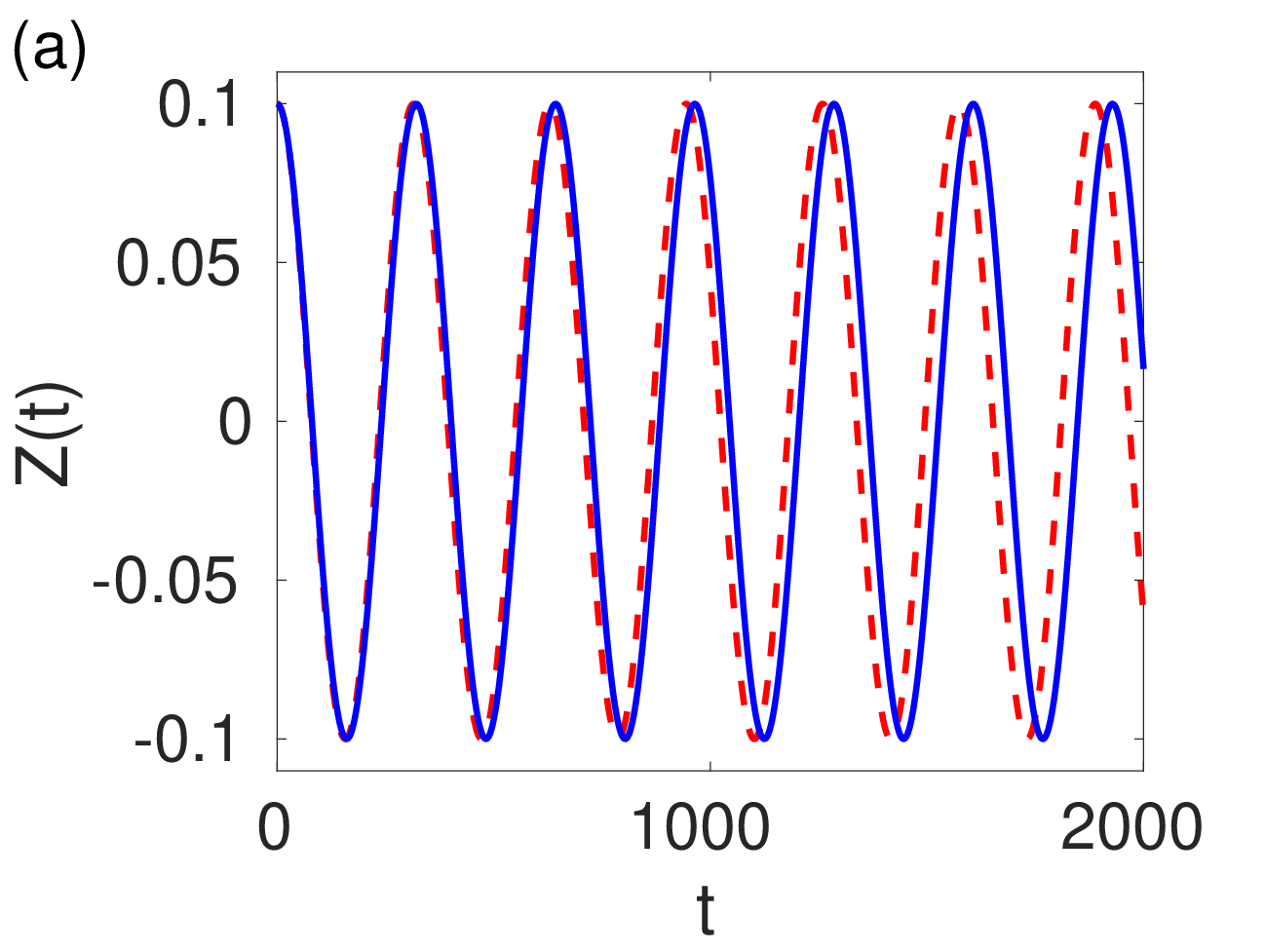}
  \includegraphics[width=4.5cm]{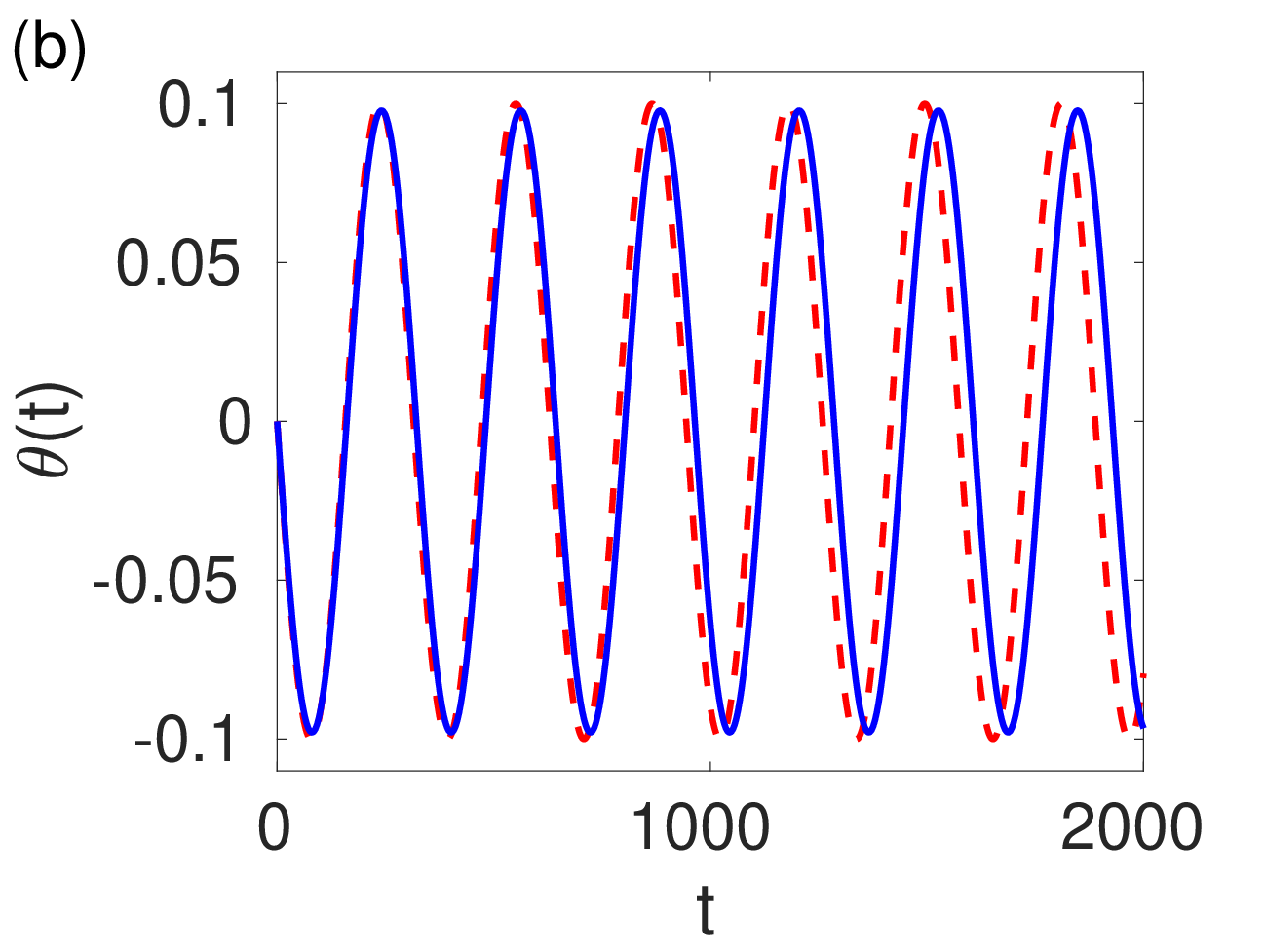} } 
\caption{The figure illustrates the time evolution of the atomic population imbalance $Z(t)$ (a) and the relative phase $\theta(t)$ (b) in the zero-phase mode. The solid curve represents the numerical simulation results, whereas the dashed curve corresponds to the theoretical predictions derived using the VA. The parameters used are $N=1$, $K=0.01$ and $g=1$.}
\label{fig-dynam}
\end{figure}

Figure \ref{fig-dynam} displays the time evolution of the atomic population imbalance $Z(t)$ and the relative phase $\theta(t)$. It can be seen from the figure that for small norms, $N=1$, the results obtained through numerical simulations and the VA are in good agreement.

% Figure 11
\begin{figure}[htbp]
  \centerline{ \includegraphics[width=4.4cm]{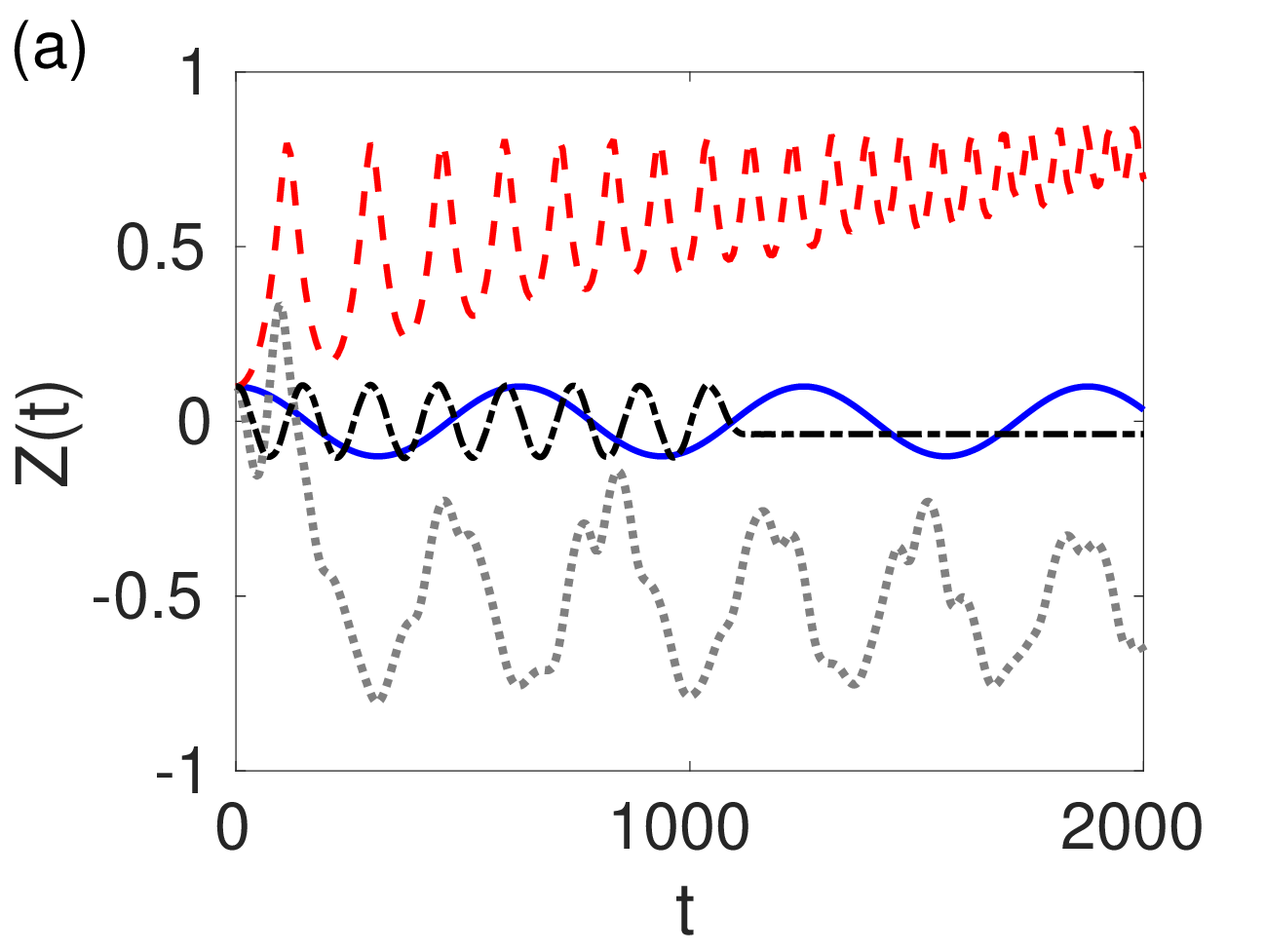}
  \includegraphics[width=4.4cm]{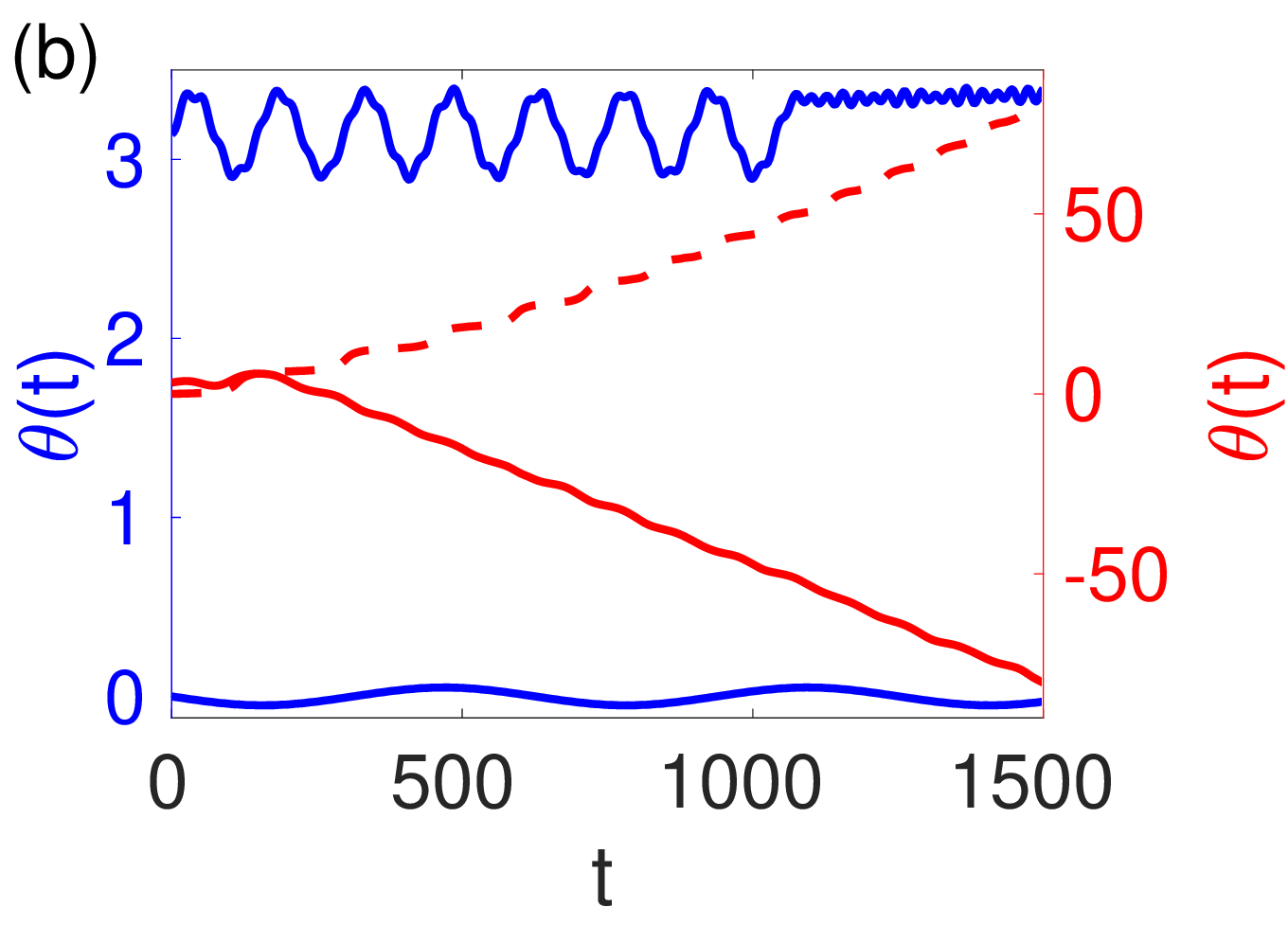}}
  \caption{(a) The time evolution of the relative imbalance and shown for the zero-phase and $\pi$-phase modes. The middle solid and the upper dashed curves correspond to the zero-phase mode for $N=100$ and $N=40$, respectively, while the middle dash-dotted and lower dotted curves represent the $\pi$-phase mode for $N=100$ and $N=5$, respectively. (b) The dynamics of the relative phase are shown for the zero-phase and $\pi$- phase modes. The upper and lower solid curves correspond to $\pi$-phase and zero-phase modes for $N=100$, respectively (left vertical axis), while the middle dashed and solid curves correspond to zero-phase mode for $N=40$ and $\pi$-phase mode for $N=5$, respectively (right vertical axis).}
\label{fig-dyn}
\end{figure}

\begin{figure}[htbp]
   \centerline{ \includegraphics[width=4.5cm]{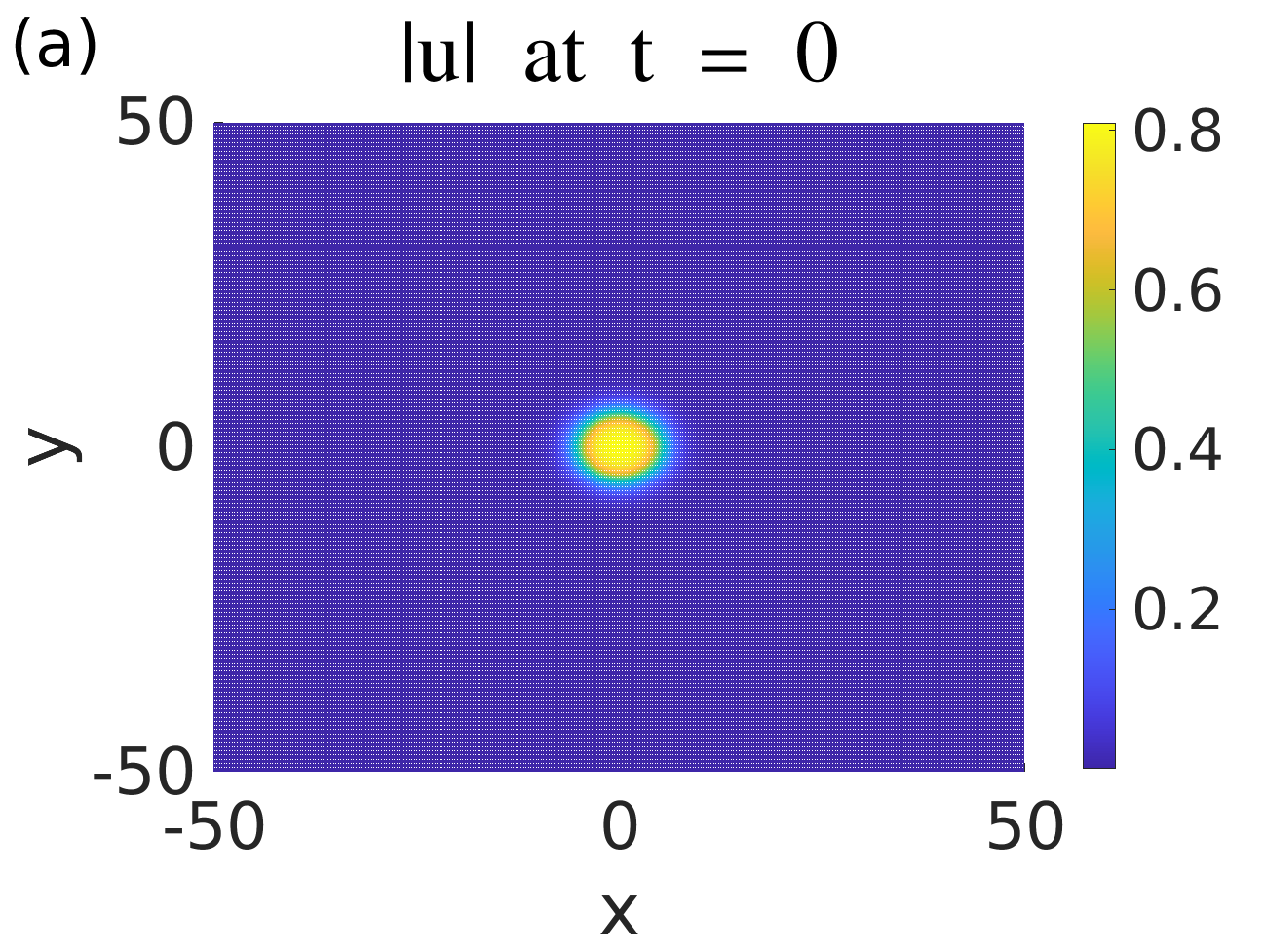} \hskip-0.1cm
      \includegraphics[width=4.5cm]{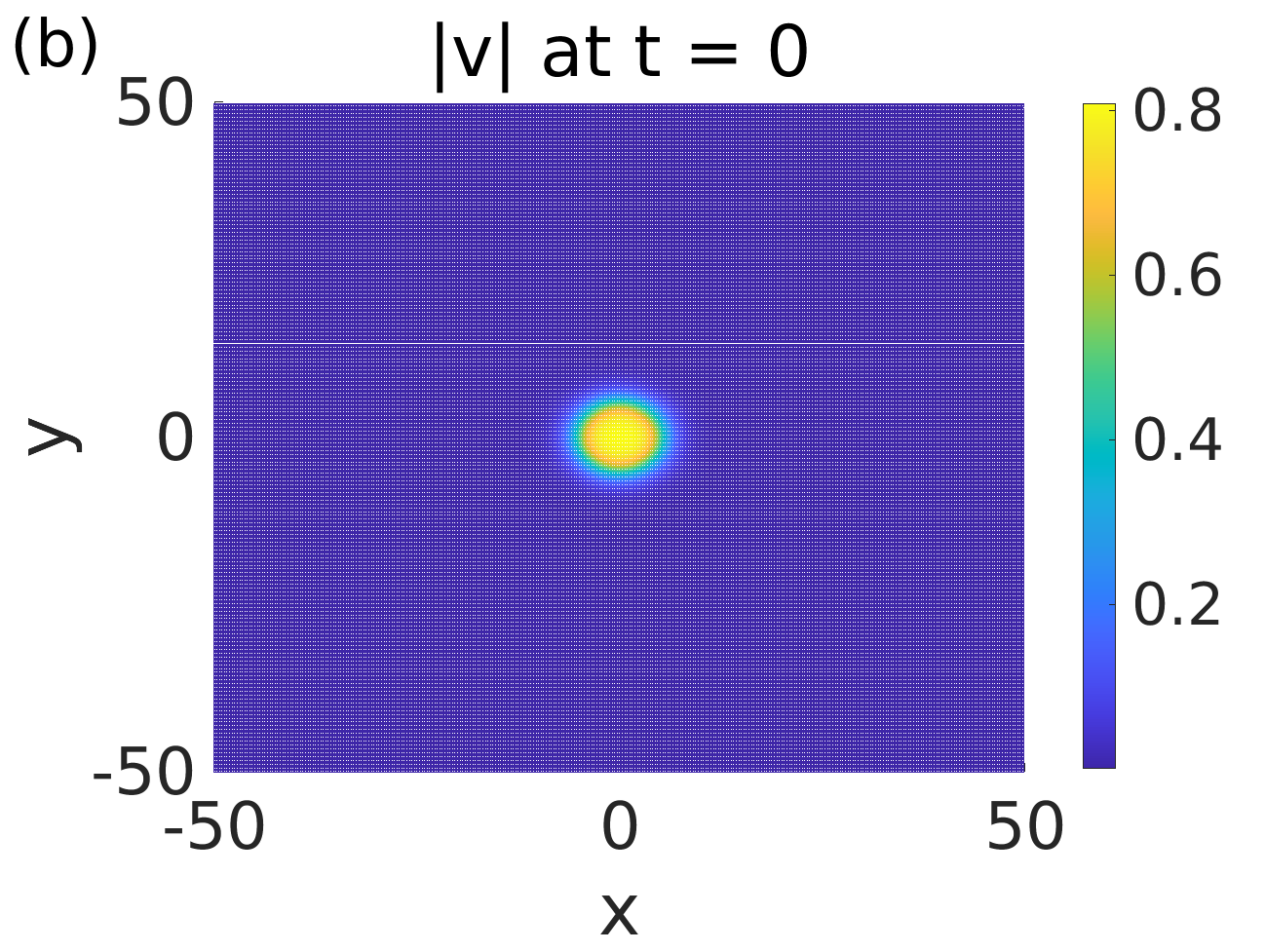}}
   \centerline{ \includegraphics[width=4.5cm]{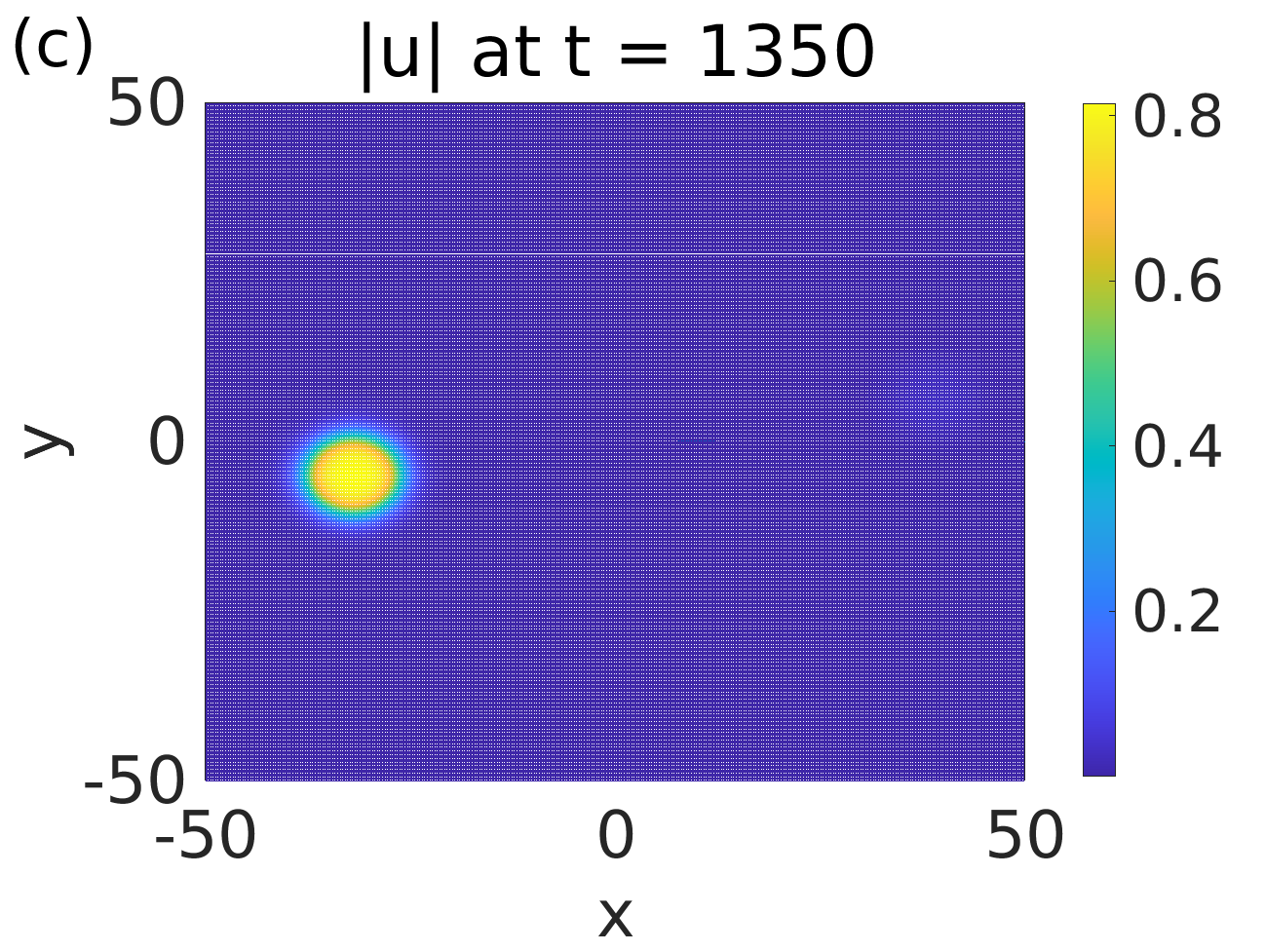} \hskip-0.1cm
      \includegraphics[width=4.5cm]{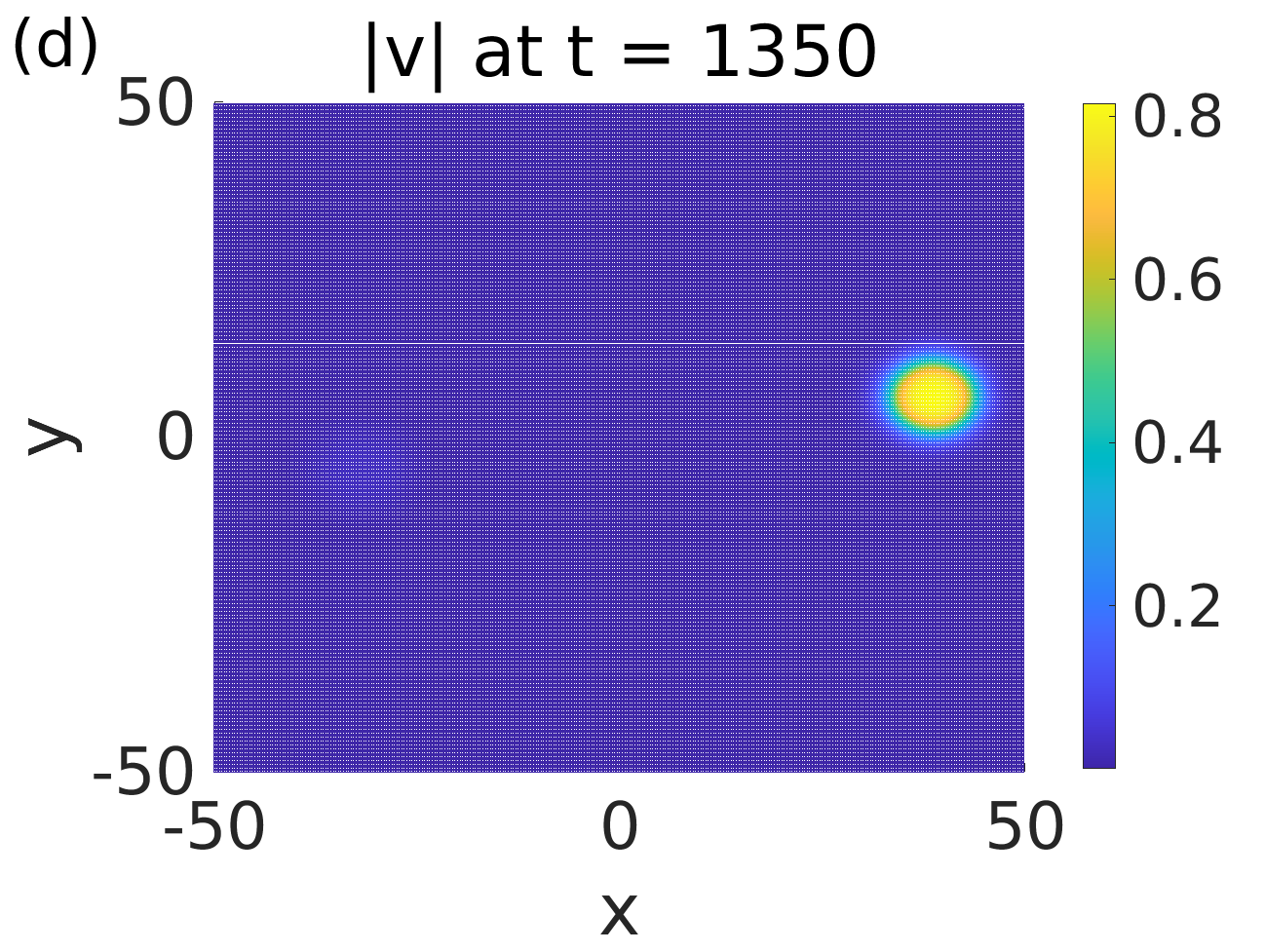}}
\caption{The figure illustrates the positions of two quantum droplets at times $t=0$ (a), (b) and $t=1350$ (c), (d) in the $\pi$-phase regime with an initial imbalance of $Z_0=0.1$. It demonstrates how the droplets move away from each other over time. Other parameters are $N=100$, $K=0.01$ and $g=1$.}
\label{fig-5abcd}
\end{figure}

In Fig.~\ref{fig-dyn}, the dynamics of the relative atomic imbalance $Z(t)$ and the relative phase $\theta(t)$ are shown for different norms and phase regimes. For $N=100$, Josephson oscillations are observed (see the middle solid and dash-dotted lines in Fig.~\ref{fig-dyn} (a)). In the $\pi$-phase regime, the droplets separate after a few Josephson oscillations, whereas in the zero-phase regime, the long-time dynamics of $Z(t)$ indicate that the droplets remain bound. The upper red dashed line and the lower gray dotted line in Fig.~\ref{fig-dyn} (a) correspond to $N=40$ and $N=5$, respectively, both representing the self-trapping regime. The associated dynamics of $\theta(t)$ are displayed in panel (b) with matching line styles: in the Josephson oscillation regime, $\theta(t)$ exhibits harmonic oscillations (left axis), while in the self-trapping regime, the relative phase increases monotonically without bound (right axis). The corresponding density profiles are shown in Figure \ref{fig-5abcd}. These observations in the dynamics of $Z(t)$ and $\theta(t)$ are further supported by the frequency spectra shown in Fig.~\ref{fig-freq}.

% Figure 12

% Figure 13
\begin{figure}[htbp]
  \centerline{ \includegraphics[width=7.55cm]{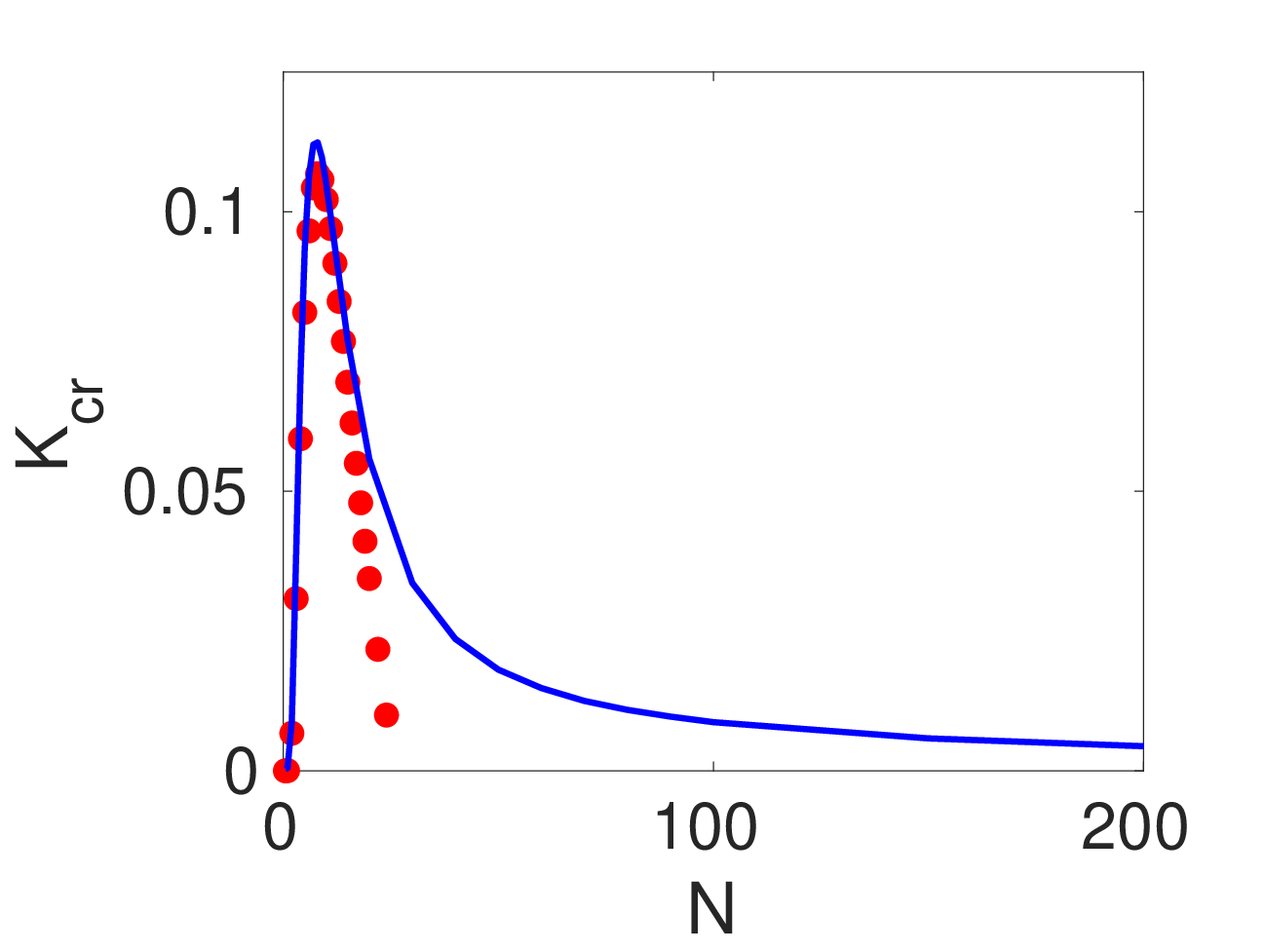}}
\caption{The figure illustrates the dependence of the critical interaction parameter $K$ on the norm $N$ for the zero-phase mode. Points represent the theoretical prediction based on Eq.~(\ref{critical_v}), while the solid line corresponds to the results of numerical simulations. The initial population imbalance is set to $Z_0=0.1$ and the quantum fluctuation parameter is $g=1$.}
\label{fig-K_c}
\end{figure}

Figure~\ref{fig-K_c} shows the dependence of the critical coupling $K_{\mathrm{cr}}$ on the norm $N$ for the zero-phase mode. The points represent the theoretical prediction from Eq.~(\ref{critical_v}), while the solid curve denotes the results of numerical simulations. For small norms ($N \lesssim 20$), the agreement is reasonable. For larger norms ($N>20$), however, a clear discrepancy emerges: the numerically obtained values of $K_{\mathrm{cr}}$ approach zero, whereas the theoretical prediction becomes negative. Hence, the two approaches agree only qualitatively. In the simulations, $K_{\mathrm{cr}}$ is determined with a numerical precision of $10^{-4}$. Notably, dynamics are self-trapped for $K>K_{\mathrm{cr}}(N)$, while Josephson oscillations occur for $K<K_{\mathrm{cr}}(N)$.

\section{Quantum droplets moving in two channels}
\label{sec:movingQDs}

In this section, we analyze moving of the QDs in two channels. There is an effective interaction energy between the QDs, which can be repulsive or attractive depending on the relative phase $\theta$. To qualitatively estimate this effective energy, we choose the solutions of Eq. (\ref{eq:coupGPE}) as simple Gaussian functions:

\begin{eqnarray}
    & u=A_1 e^{-((x-x_1)^2+ (y-y_1)^2)/2a_1^2 + i\phi_1} \nonumber \\
    & v=A_1 e^{-((x-x_2)^2+ (y-y_2)^2)/2a_2^2 + i\phi_2}
\end{eqnarray}
Then the energy of interaction can be calculated from the overlap integral
$$
E_{int} = K\int\int dx dy (u^{\ast}v + v^{\ast}u)
$$
Substituting the expressions for the trial functions $u_i$ and calculating integrals, we obtain
\begin{equation}\label{eq:int}
E_{int} = A_1A_2 \frac{\pi}{a_1^2 + a_2^2}e^{-\alpha^2 \Delta R^2}\cos(\theta),
\end{equation}
where
$$
 \alpha^2 = \frac{a_1^2a_2^2}{2(a_1^2 + a_2^2 )},
\Delta R^2 = (x_1-x_2)^2 + (y_1-y_2)^2, \theta =\phi_1 - \phi_2.
$$
The character of interaction between droplets depends on the value of their relative phase difference $\theta$.  For $\theta =0$, we have the attractive interaction; for $\theta =\pi$, the interaction is repulsive. 

\begin{figure}[htbp]
   \centerline{ \includegraphics[width=4.4cm]{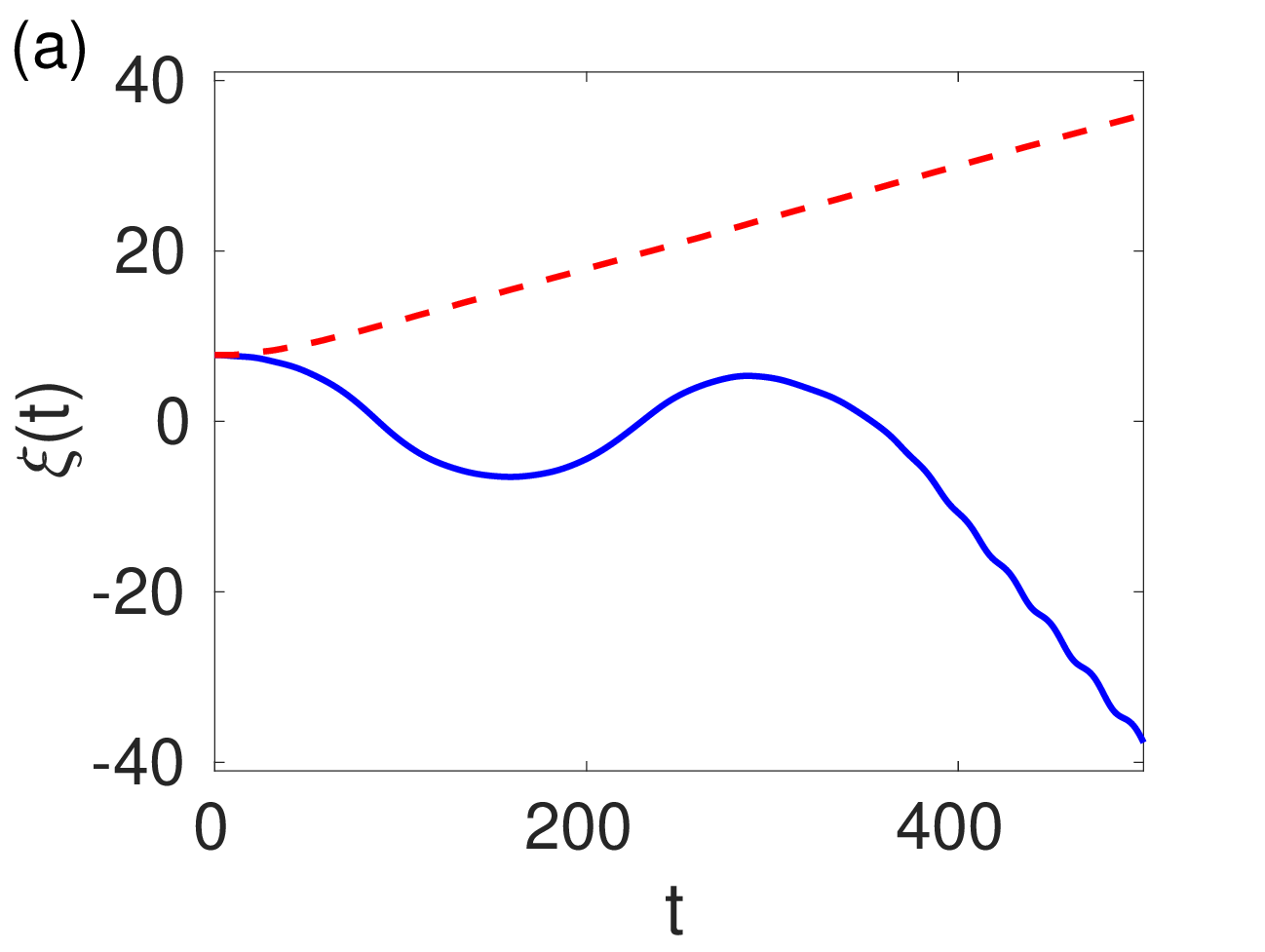} \hskip-0.1cm
      \includegraphics[width=4.7cm]{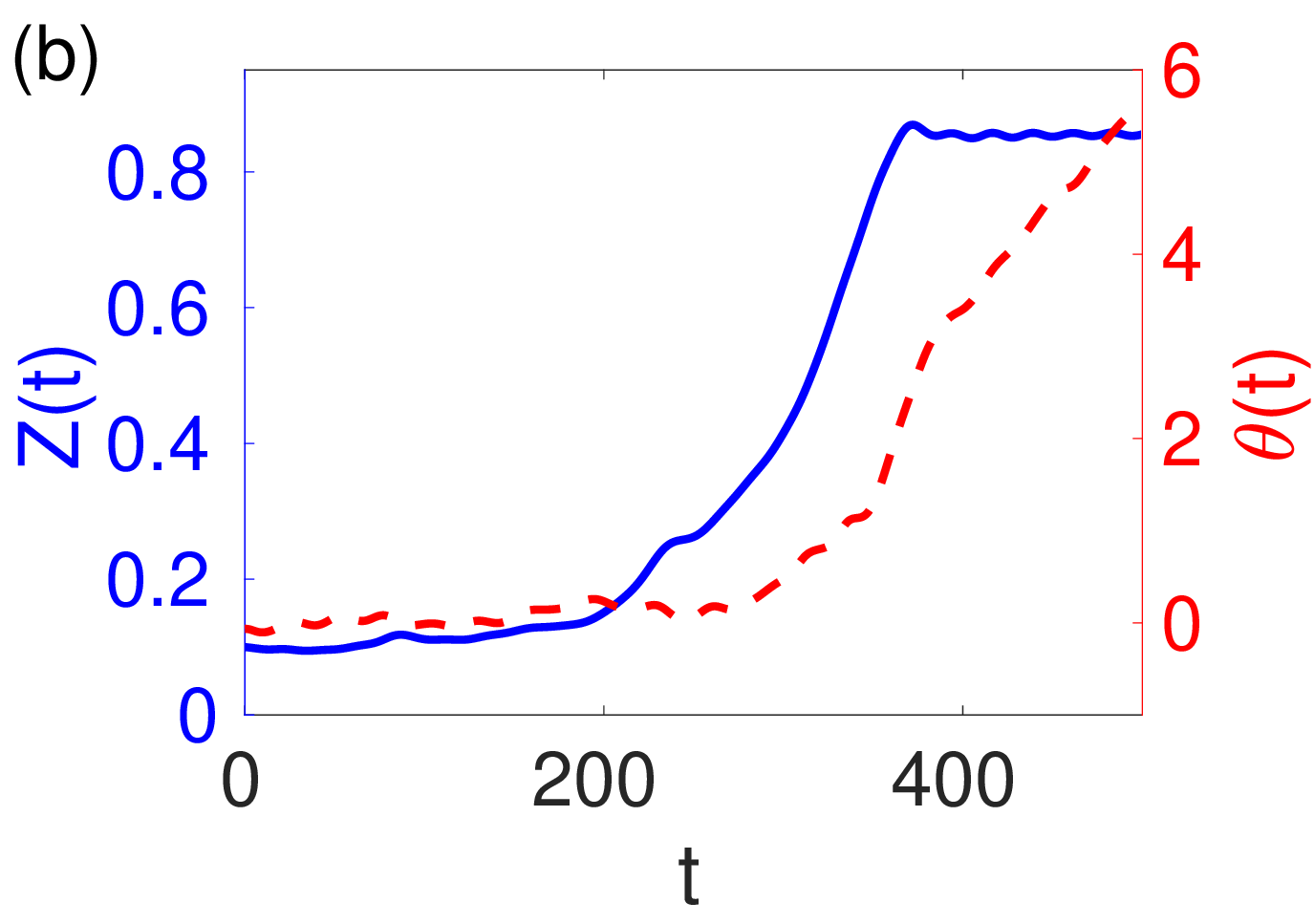}}
   \centerline{ \includegraphics[width=4.4cm]{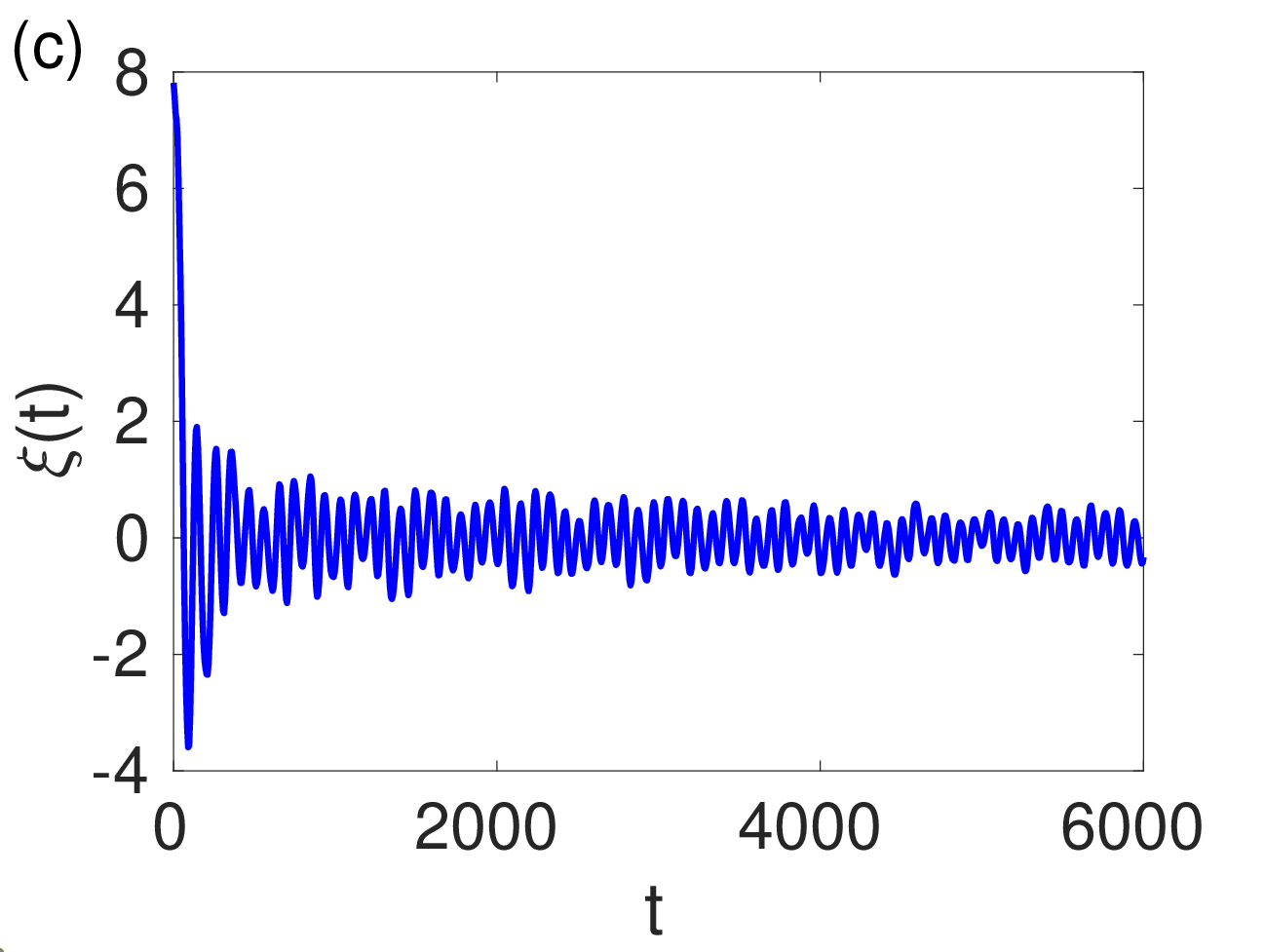} \hskip-0.1cm
      \includegraphics[width=4.7cm]{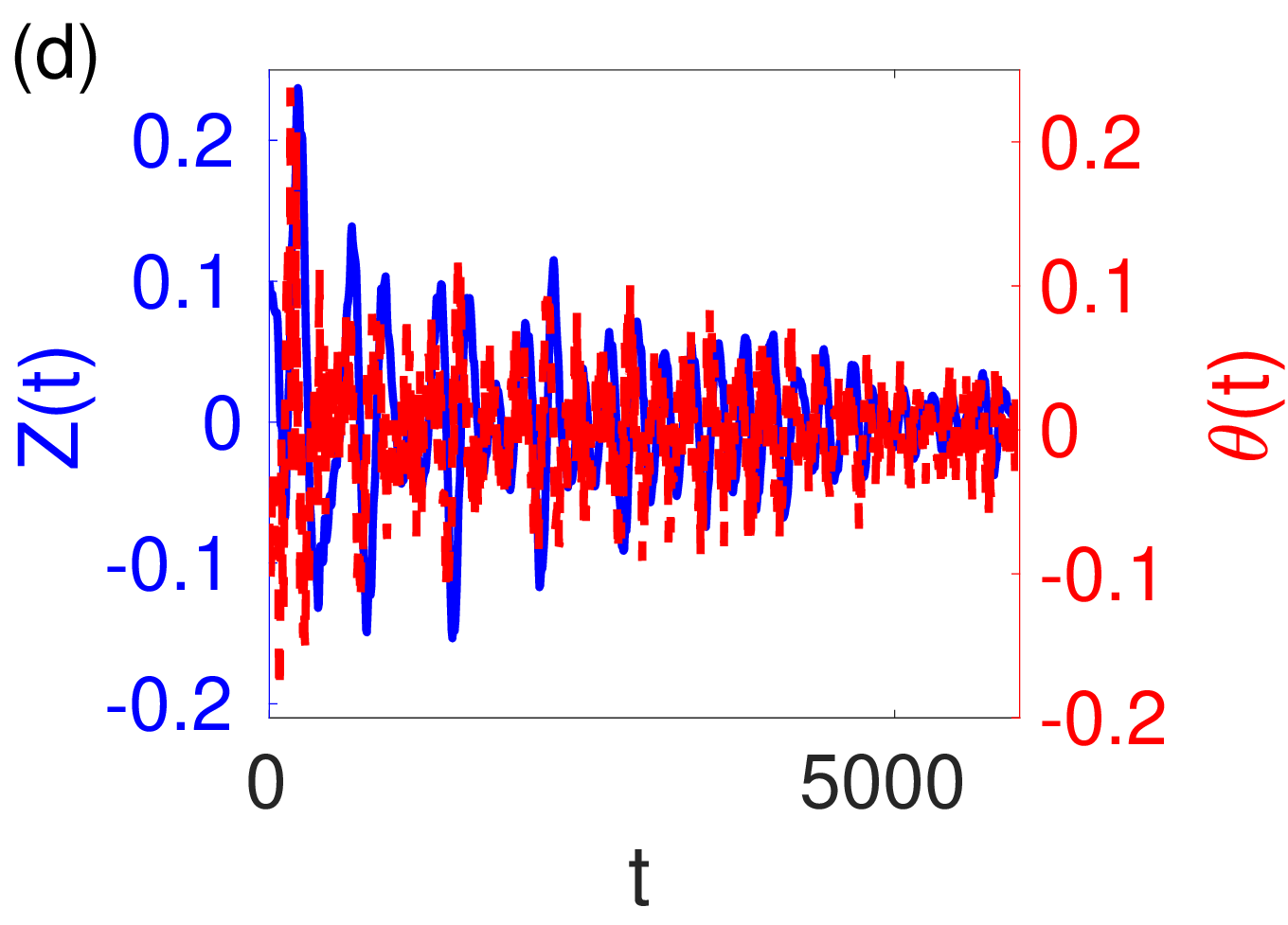}}
\caption{Time evolutions for two droplets launched head-on from symmetric channels. Panels (a,c) show the relative centre of mass displacement, $\xi(t)=x_2-x_1$, while panels (b,d) show the atom number imbalance $Z(t)$ (solid, left axis) and the relative phase $\theta(t)$ (dashed, right axis). In (a), the solid line denotes the zero-phase and the dashed line the $\pi$-phase. Panels (b--d) display only the zero-phase because, in the $\pi$-phase, the droplets separate without oscillations (see panel (a)). Parameters: $K=0.01$ for (a,b) and $K=0.02$ for (c,d). Other fixed initial parameters are $x_2=-x_1\approx3.91$, $y_2=-y_1\approx3.91$, $Z_0=0.1$, $N=100$, and $g=1$. }
\label{fig-moving}
\end{figure}

Although the expression given in Eq. (\ref{eq:int}) is formally valid only for small values of $N$, it qualitatively captures the behaviour of the interaction energy even for larger $N$. Figure \ref{fig-moving} (a) illustrates the time evolution of the relative displacement of the centre of mass between two droplets initially confined to separate channels. The solid curve corresponds to the zero-phase regime, while the dashed curve represents the $\pi$-phase regime. The displacement is shown only along the $x$-axis; however, a similar dynamical behaviour is observed along the $y$-axis. As evident from the figure, in the zero-phase regime, the droplets undergo a brief oscillation before moving apart. This behaviour is attributed to the evolution of the relative phase $\theta(t)$. Since $\theta(t)$ is non-linearly coupled to the atomic imbalance $Z(t)$ (see Eq.(\ref{eq:QD_JosepEq})), variations in $Z(t)$ lead to the corresponding changes in $\theta(t)$, as shown in Fig. \ref{fig-moving} (b). As $\theta(t)$ increases, the interaction energy decreases according to Eq. (\ref{eq:int}), which weakens the attractive force between the droplets and causes them to separate. In contrast, in the $\pi$-phase regime, the initial interaction is repulsive (see Eq. (\ref{eq:int})), and the droplets immediately repel each other without exhibiting oscillatory motion (see Fig. \ref{fig-moving} (a) dashed line). A similar behaviour is observed for the initial relative phase of $0.25\pi$ and $0.5\pi$. Figures \ref{fig-moving} (a) and (b) correspond to the case where the interaction parameter is set to $K = 0.01$, under which no Josephson oscillations are observed.

Figures \ref{fig-moving} (c) and (d) correspond to the case with $K = 0.02$ and are presented for the zero-phase mode. In this case, the interaction between the droplets is stronger, and since the relative phase oscillates around zero (see Fig. \ref{fig-moving} (d)), the centres of mass of the droplets exhibit oscillations relative to each other over time. As a result, Josephson transitions become apparent (see Fig. \ref{fig-moving} (d)). However, because the distance between the droplets oscillates in time, the resulting Josephson oscillations are not perfectly harmonic.

% Figure 14

\begin{figure}[htbp]
\centerline{%
\includegraphics[width=4.5cm]{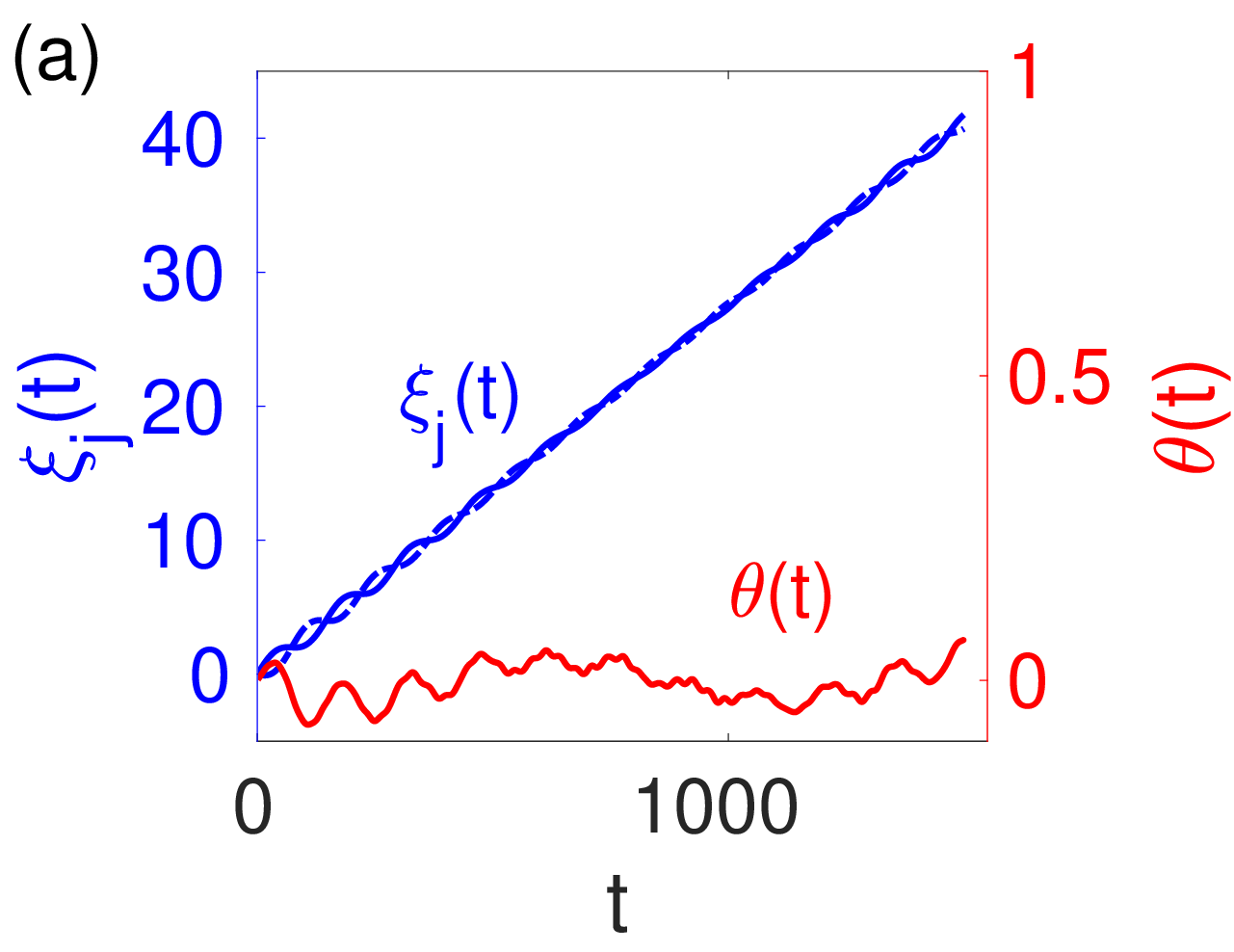}\hskip 0.05cm
\includegraphics[width=4.5cm]{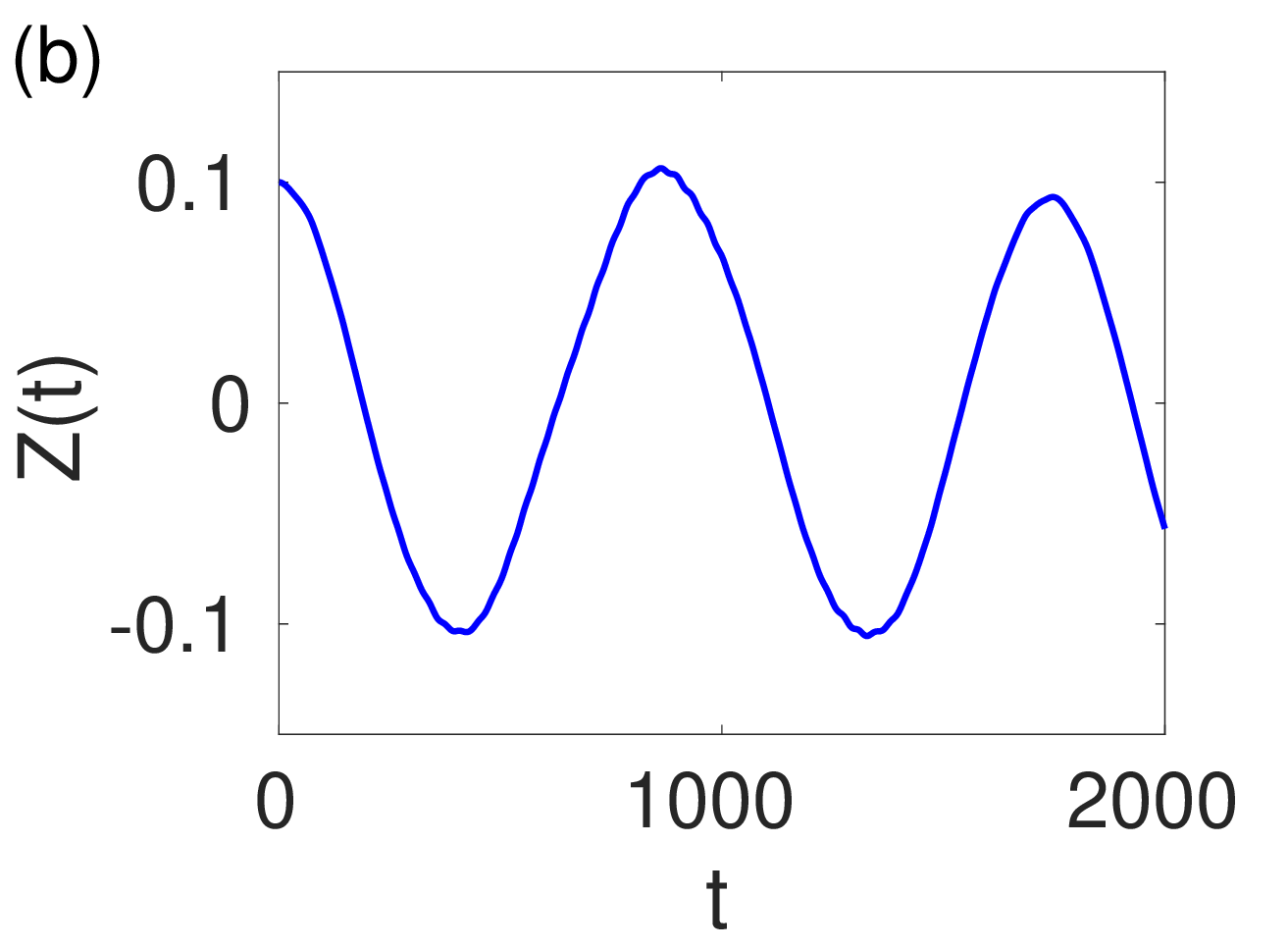}}
\caption{(a) Time evolution of the center-of-mass coordinates $\xi_j(t)$ (right axis) and the relative phase $\theta(t)$ (left axis). The solid blue and dash-dotted blue curves depict $\xi_j(t)$ for components $j=1,2$, respectively; the solid red curve shows $\theta(t)$. (b) Dynamics of the population imbalance $Z(t)$ between the droplets. Parameters: $\theta_0=0$, $g=1$, $N=100$, and $K=0.02$.}
\label{fig-AB}
\end{figure}

The Andreev-Bashkin effect refers to a nondissipative (entrainment) drag between two superfluids, where a superflow in one component induces a mass current in the other~\cite{Andreev1976,Nespolo2017}. In our system, the Andreev-Bashkin effect may also occur: if one quantum droplet is set into motion, under certain conditions, it can drag the second droplet, leading to their joint motion. The Andreev-Bashkin effect is illustrated in Figure~\ref{fig-AB}, where the flow of one quantum fluid induces another to flow in the same direction without viscosity. At the initial time ($t=0$), the droplet of the second component is given a small kick (phase gradient $k=0.02$ along the positive $x$-axis. During the subsequent evolution, the kicked droplet drags the first-component droplet, which has no initial velocity, so that both move together. The time evolutions of their centres of mass and the relative phase are shown in Fig.~\ref{fig-AB}(a).

% Figure 15

While moving together, the droplets also undergo Josephson tunnelling between components, as seen in Fig.~\ref{fig-AB}(b). This behaviour is consistent with the presence of the Andreev-Bashkin effect in our system.

\section{Conclusion}
We investigate a binary Bose gas in a symmetric, dual-core, pancake-shaped trap. The system is modelled by two linearly coupled two-dimensional Gross-Pitaevskii equations that include Lee-Huang-Yang corrections. We begin with a spatially uniform condensate, derive the Hamiltonian, and obtain the Josephson oscillation frequencies in the zero- and $\pi$-phase modes. For each phase, we analyse the phase portrait, distinguishing finite and infinite trajectories and the separatrix that divides them. Finite trajectories correspond to Josephson oscillations, while infinite ones reflect self-localisation. For both the zero- and $\pi$-phase modes we derive, in closed form, the critical linear coupling (or, equivalently, the critical initial imbalance) that places the system on the separatrix. We also determine the parameter conditions set by the strength of quantum fluctuation interactions and the linear coupling under which bifurcations occur. In the zero-phase, two pitchfork bifurcations appear in sequence: a supercritical bifurcation at smaller norms followed by a subcritical one at larger norms. The resulting bifurcation diagram exhibits bistability and hysteresis. In the $\pi$-phase, we find a single subcritical pitchfork. 

We then turn to quantum droplets and develop a variational approach for their coupled dynamics. Stationary droplet parameters are obtained via the variational method and imaginary-time simulations, and we verify their stability in direct numerical simulations. The VA predicted profiles match the numerically computed QDs with high accuracy. We also derive analytic expressions for the Josephson frequencies in both the zero- and $\pi$-phase droplet regimes and confirmed these against numerical simulations. For smaller particle numbers, the variational approach results for the Josephson oscillations agree well with numerics. For larger particle numbers, the values of the predicted frequencies can differ by orders of magnitude, but the qualitative trends remain consistent, so we rely primarily on numerical simulations in that regime.

In the $\pi$ phase for quantum droplet tunnelling, we observe that the two droplets split in opposite directions after roughly $6-7$ oscillation periods. This splitting is explained by evaluating the droplet-droplet interaction energies. In numerical simulations, we map the critical linear coupling that separates Josephson and self-trapped regimes as the particle number varies. Finally, in the two-core geometry, we examine inter-droplet interactions and find numerical signatures consistent with the Andreev-Bashkin superfluid drag effect.

\section{Acknowledgments}
This work has been supported from the State Budget of the Republic of Uzbekistan (Grant No. 2025 year award).

\end{document}